\documentclass[twocolumn]{aastex63}

\usepackage{hyperref}
\let\oldvec\vec
\usepackage{amsmath}

\shortauthors{Paine et al.}

\begin{document}

\title{ Secular Extragalactic Parallax: Measurement Methods and Predictions for Gaia}

\correspondingauthor{Jennie Paine}
\email{Jennie.Paine@colorado.edu}

\author{Jennie Paine}
\affiliation{Center for Astrophysics and Space Astronomy, 
Department of Astrophysical and Planetary Sciences, 
University of Colorado, 389 UCB, Boulder, CO 80309-0389, USA}

\author{Jeremy Darling}
\affiliation{Center for Astrophysics and Space Astronomy, 
Department of Astrophysical and Planetary Sciences, 
University of Colorado, 389 UCB, Boulder, CO 80309-0389, USA}

\author{Romain Graziani}
\affiliation{University of Lyon, UCB Lyon 1, CNRS/IN2P3, IP2I Lyon, France}
\affiliation{Universit\'e Clermont Auvergne, CNRS/IN2P3, Laboratoire de Physique de Clermont, 
F-63000 Clermont-Ferrand, France }

\author{H\`{e}l\'{e}ne M. Courtois}
\affiliation{University of Lyon, UCB Lyon 1, CNRS/IN2P3, IP2I Lyon, France}

\begin{abstract}
Secular extragalactic parallax caused by the solar system's velocity relative to the cosmic microwave 
background rest frame may be observable as a dipole proper motion field with amplitude $78~\mu \rm{as~yr}^{-1}$~Mpc. 
Nearby galaxies also exhibit proper motions caused by their transverse peculiar velocities 
that prevent detection of secular parallax for any single galaxy, although a statistical
detection may be made instead. 
Such a detection could constrain the local Hubble parameter. 
We present methods to measure secular parallax using correlated extragalactic proper motions 
and find a first limit on the secular parallax amplitude using proper motions of 232 nearby galaxies from
{\it Gaia} Data Release 2. The recovered dipole has insignificant upper limit of 3500~$\mu$as~yr$^{-1}$~Mpc.
This measurement will be improved by larger sample size and reduced proper motion uncertainties in future 
data releases. 
Using the local peculiar velocity field derived from Cosmicflows-3, we simulate galaxy proper motions and
predict that a significant detection ($5-10\sigma$) of the secular parallax amplitude will be possible by {\it Gaia}'s end of mission.
The detection is contingent on proper motions of nearby ($<5$~Mpc), bright ($G<15$~mag) galaxies, 
and corresponds to an insignificant upper limit on the Hubble parameter.
We further investigate the implications of our simulations for the study of transverse peculiar velocities, 
which we find to be consistent with large scale structure theory. 
The peculiar velocity field additionally results in low-multipole correlated proper motions on the order of $0.3~\mu$as yr$^{-1}$
that may be confounded with other cosmological proper motion measurements, 
such as limits on the gravitational wave background and the anisotropy of the Hubble expansion. 
\end{abstract}

\keywords{Astrometry  --- Proper motions --- Astronomical methods --- Observational cosmology --- Large-scale structure of the universe}

\section{Introduction}\label{sec:intro}

A galaxy's proper motion reflects a combination of  its peculiar velocity, cosmological effects, 
and observer induced apparent motions. One such observer induced
motion is the secular extragalactic parallax caused by the solar system's velocity with
respect to the cosmic microwave background (CMB) rest frame
\citep{Kardashev1986, Ding2009, Bachchan2016}. 
The CMB temperature dipole 
has an amplitude of about 369 km s$^{-1}$ toward the direction $(l,b) = (264^\circ, 48^\circ)$,
$(168^\circ, -7^\circ)$ in RA and Dec., 
and corresponds to a linear solar velocity of about 78 AU yr$^{-1}$ \citep{Hinshaw2009}. 
This velocity is a combination of the solar system's orbit in the Galaxy and the Galactic 
peculiar motion and will therefore cause a parallax shift away from the direction of motion for extragalactic objects
that is distinct from the annual parallax caused by the Earth's orbit in the solar system.
The resulting secular parallax may be observed as a proper motion of amplitude 
$78 ~ \mu \rm{as~yr}^{-1} (\frac{1 Mpc}{D}) |\sin{\beta}|$, 
where $D$ is the proper motion distance of the galaxy, which is equivalent to the
comoving distance for a flat universe \citep{Hogg1999}, 
and $\beta$ is the angle between the position of the galaxy and the CMB apex. 
The global signal is therefore a proper motion dipole that diminishes with distance. 

Detection of secular parallax for any individual galaxy is complicated by confounding 
proper motions. The largest expected contributions to the proper motion are 
the peculiar velocities caused by gravitational interactions 
with large-scale structure (LSS) and the secular aberration drift caused by 
the solar system barycenter acceleration about the Galactic center. 
Secular aberration drift is observed as a distance-independent
dipole with amplitude of $\sim5~\mu$as yr$^{-1}$ \citep{Titov2013, Truebenbach2017}, 
which should be separable from a distance dependent dipole.
LSS, however, induces distance-dependent proper motions at all 
angular scales with amplitudes comparable to secular parallax \citep{Hall2018}. 
For individual galaxies, separating the peculiar and secular parallax components of 
the proper motion is not possible without independent estimates of the peculiar velocity 
and distance of the galaxy, though the inferred secular parallax distance would not 
provide an independent distance estimate in this case. 
However, a statistical detection of secular parallax may be possible for large 
sample of galaxy proper motions, which may be used to calibrate existing extragalactic
distance measures and to constrain the Hubble parameter \citep{Ding2009, Hall2018}.

To detect the global secular parallax dipole, one needs a large sample of nearby
galaxies with proper motion measurements. 
The {\it Gaia} mission's Data Release 2 (DR2) contains proper motions 
for at least a half million extragalactic objects identified as mid-infrared 
active galactic nuclei (AGN; \citealt{Lindegren2018}). 
Most {\it Gaia} AGN are either too distant to be useful for secular parallax or may 
not have distance measures, so in this work we target a new sample of more 
local galaxies in {\it Gaia} DR2 using the Cosmicflows-3 catalog \citep{Tully2016}. 
Individual galaxies should not have significant proper motion measurements from {\it Gaia}, 
but with large sample size and even sky distribution, we may constrain correlated 
proper motions with smaller amplitudes than the individual uncertainties. 
Future data releases should increase the number of extragalactic objects with measured 
proper motions and should have substantially lower uncertainties. 

The expected secular parallax, peculiar proper motions, and higher multipole signals 
are detailed in Section~\ref{sec:methods}.  
We present a first secular parallax limit using {\it Gaia} DR2 proper motions for a sample
of nearby Cosmicflows-3 galaxies in Section~\ref{sec:dr2}.
In Section~\ref{sec:predictions}, we utilize the Cosmicflows peculiar velocity field 
to simulate galaxy proper motions consistent with {\it Gaia's} end-of-mission performance
and predict the detection of secular parallax.
We analyze the peculiar proper motions in detail in Section~\ref{sec:peculiar}.
Discussion and main conclusions are given in Section~\ref{sec:discussion}. 
We assume a flat cosmology and a Hubble constant of $H_0=70$ km s$^{-1}$ Mpc$^{-1}$, 
which is $H_0 = 15 ~\mu$as yr$^{-1}$ in proper motion units. 

\section{Characterizing the Proper Motion Vector Field } \label{sec:methods}

\begin{figure}[t!]
    \centering
    \includegraphics[width=\columnwidth]{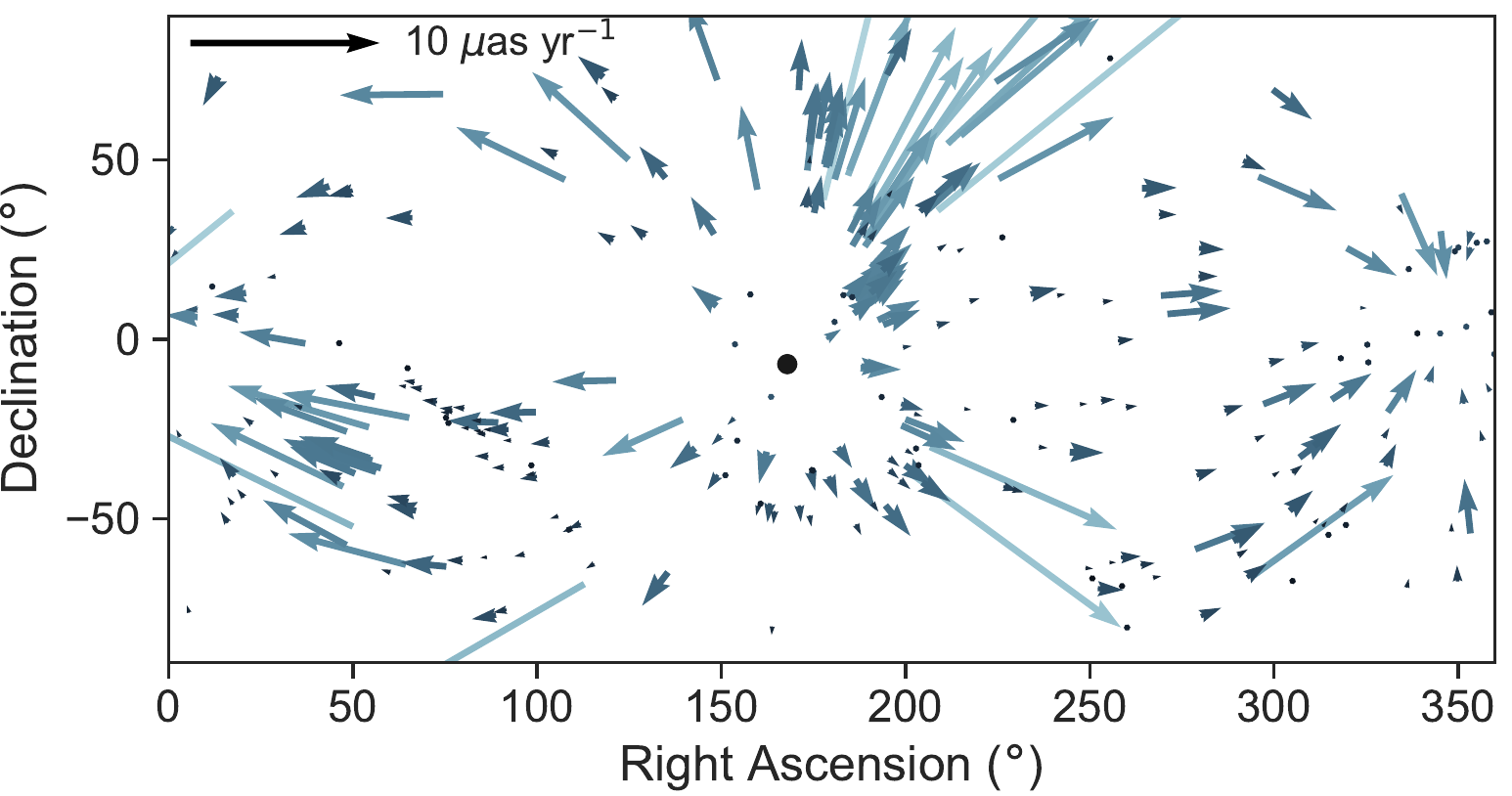}
    \caption{The expected secular parallax proper motions for 232 nearby {\it Gaia} galaxies, which 
    are described in Section~\ref{sec:dr2_selection}. 
    The colors of the arrows scale with distance, which range from $~3-410$ Mpc. 
	The direction of the solar system motion with respect to the CMB is indicated by the black dot. }
    \label{fig:parallax_expected}
\end{figure}

The term ``secular parallax" may have two definitions. 
The first usage refers to the phenomenon of
an apparent distance-dependent dipolar proper motion field caused by the observer's velocity relative to the bulk motion
of a sample of objects. 
In this work, we use ``secular parallax" to describe the proper motion dipole
predicted from the solar system's velocity measured with respect to the CMB reference frame, 
which may be leveraged to constrain galaxy distances.
The former definition includes the dipole correlated components of galaxy velocities with respect
to the CMB frame. We distinguish such peculiar proper motions from 
secular parallax proper motions, although the net dipole one measures is the sum of the two effects. 
Secular parallax detection as a method to measure galaxy distances or the Hubble
constant is contingent on knowing the underlying velocity as a prior, and therefore one must separate 
the secular parallax from the peculiar transverse motions that are not as well constrained. 
In this section, we describe the expected secular parallax and other relevant proper motion components.

The global secular parallax field is modeled as a distance-dependent E-mode (curl-free) dipole
where galaxies appear to stream away from the direction of motion \citep{Ding2009}. 
For example, a  galaxy at 1 Mpc and $90^\circ$ from the CMB apex will have a secular parallax, 
observed as a proper motion, of $78 ~ \mu \rm{as~yr}^{-1}$ oriented away from the direction of 
the CMB dipole apex. The magnitude of the proper motion is modulated by the angle between 
the galaxy and the direction of the apex, $\beta$ \citep{Ding2009}, given by the following expression:
\begin{equation}
  | \mu |  = \left(78 ~\mu \rm{as~yr}^{-1}~\rm{Mpc}\right) \left( \frac{1 }{D} \right) |\sin\beta|. 
  \label{eqn:parallax}
\end{equation}
Figure~\ref{fig:parallax_expected} shows the expected parallax dipole proper motions 
for a sample of {\it Gaia} galaxies described in Section~\ref{sec:dr2_selection} which we 
use to limit the secular parallax signal. 

Following the notation of \cite{Mignard2012}, an E-mode dipole vector
field may be expressed as an $\ell =1$ vector spherical harmonic:
\begin{align}
  \mathbf{\oldvec{V}}_{E1} (\alpha,\delta) &= 
      \sqrt{3\over 4\pi} \left(s_{11}^{Re}\,  \sin\alpha+s_{11}^{Im}\,  \cos\alpha\right) \mathbf{\hat{e}}_\alpha \notag \\
   &  + \sqrt{3\over 4\pi}  \Big(s_{10} \sqrt{1\over 2}\, \cos\delta + s_{11}^{Re}\, \cos\alpha\sin\delta \label{eqn:dipole}\\
  & ~~~~~~~~~~~~~~-s_{11}^{Im}\,  \sin\alpha\sin\delta\Big) \mathbf{\hat{e}}_\delta \notag
\end{align}
where the coefficients $s_{10}$ and $s_{11}^{Re,Im}$ determine the direction and amplitude
of the vector field, and $\alpha$ and $\delta$ are the coordinates in RA and Dec. 
The vectors $\mathbf{\hat{e}}_\alpha$ and $\mathbf{\hat{e}}_\delta$ are unit vectors in RA and Dec.
The power of the vector field is the integral of the squared field over the unit sphere and is calculated from the coefficients by
\begin{equation}
P_{E1} = s_{10}^2 + 2 \left((s_{11}^{Re})^2 + (s_{11}^{Im})^2 \right).
\label{eqn:power}
\end{equation}
The power is related to the dipole amplitude by
\begin{equation}
A_{E1} =\sqrt{{3\over 8\pi} P_{E1}},
\label{eqn:amplitude}
\end{equation}
where the factor of $\sqrt{3/8\pi}$ comes from integrating over $4\pi$ sr.

The distance dependence of secular parallax may be incorporated by expressing the 
coefficients  in units of $\mu$as yr$^{-1}$ Mpc (velocity units).
The solar system's velocity with respect to the CMB is $369 \pm0.9 \rm{~km~s}^{-1}$
in the direction $(l,b) = (264^\circ, 48^\circ)$ \citep{Hinshaw2009}, 
which causes an observed proper motion dipole with coefficients 
$(s_{10}, s_{11}^{Re}, s_{11}^{Im}) =$ $(27.2,-155.0, -33.1)~\mu\rm{as~yr}^{-1}$~Mpc. 
The secular parallax dipole root power is $\sqrt{P_{E1}}=226~\mu$as yr$^{-1}$~Mpc
and the amplitude is 78 $\mu$as yr$^{-1}$~Mpc, as given by Equation~\ref{eqn:parallax}.

Two other E-mode dipolar proper motions of interest in this work are the secular aberration 
drift and the peculiar proper motion dipole caused by LSS. The former is 
a distance-independent dipole that has been previously detected using quasar proper motions
from very long baseline interferometry \citep{Titov2013,Truebenbach2017}.
The secular aberration drift amplitude is $\sim 5 \mu$as yr$^{-1}$, which will dominate dipolar 
proper motions for galaxies at distances of approximately 16 Mpc or greater. 
We expect that the secular aberration drift
will be detected with {\it Gaia} using high redshift AGN \citep{Paine2018}, enabling the 
study of distance-dependent dipoles after the distance-independent signal is subtracted. 

Peculiar velocities induce
distance-dependent proper motions at all angular scales, including a dominant dipole
for local galaxies. \cite{Hall2018} presented predictions for the LSS transverse velocity 
power spectrum and its correlation to the secular parallax dipole. The transverse peculiar 
velocity dipole causes proper motions of similar amplitude to secular parallax for galaxies 
within $\sim 100$ Mpc. \cite{Hall2018} therefore predict transverse peculiar velocities 
dominate the error on secular parallax measurements for local galaxies. The predicted 
peculiar velocity dipole decreases in power as a function of distance, which causes the observed peculiar 
proper motion dipole amplitude to decrease faster than $1/D$. 
The distance dependence of peculiar proper motions means that the peculiar dipole may not
be fit separately and subtracted prior to fitting for secular parallax, unlike the secular 
aberration drift. 

In Section~\ref{sec:peculiar}, we investigate higher multipole vector fields associated 
with peculiar proper motions. Expressions for $\ell=2,3$ vector fields may be found in 
\cite{Mignard2012}. The power for any general $\ell$ is 
\begin{equation}
	P_l = s^2_{\ell 0} + t^2_{\ell 0}  + 2 \sum_{m=1}^\ell \left((s_{\ell m}^{Re})^2 + (s_{\ell m}^{Im})^2 + (t_{\ell m}^{Re})^2 + (t_{\ell m}^{Im})^2  \right),  
\end{equation}
where $s$ and $t$ denote the spheroidal (E-mode) and toroidal (B-mode) components 
of the vector field, respectively. 

The total proper motion field is then the vector sum of distance-dependent and distance-independent components:
\begin{align}
 \mathbf{\oldvec{\mu}} (D) =  \mathbf{\oldvec{V}}_{E1,SP} ~D^{-1}  + \mathbf{\oldvec{V}}_{E1,PV}  ~D^{-1}\\
       + \sum_{\ell=2}^{\infty} \left(\mathbf{\oldvec{V}}_{E\ell,PV} ~D^{-1}  + \mathbf{\oldvec{V}}_{\ell}\right), \notag
\label{eqn:total_field}
\end{align}
where $ \mathbf{\oldvec{V}}_{E1,SP}$ denotes the dipole induced by the solar system velocity with respect to the CMB. 
$\mathbf{\oldvec{V}}_{E1,PV}$ and $\mathbf{\oldvec{V}}_{E\ell,PV}$
denote the dipole and higher order E-modes induced by peculiar velocities. 
The secular aberration drift and higher order
distance-independent modes are represented by $\mathbf{\oldvec{V}}_{\ell}$.  
Note that the secular parallax and peculiar components have units of velocity, $\mu$as yr$^{-1}$ Mpc, 
whereas the distance-independent multipoles are given in units of proper motion, $\mu$as yr$^{-1}$.

\section{A First Secular Parallax Limit} \label{sec:dr2}

\subsection{Sample Selection}\label{sec:dr2_selection}

Making a statistical detection of extragalactic parallax will require a large sample of
nearby galaxies with both proper motion and distance measurements.
For this purpose, we selected galaxies from  Cosmicflows-3 \citep{Tully2016}, 
a catalog of 17,669  redshift-independent galaxy distances. The majority of distances were
measured using either the relation between galaxy rotation and luminosity 
(the Tully-Fisher relation; \citealt{Tully1977}) or the relations between the velocity dispersion,
radius, and luminosity of elliptical galaxies (the fundamental plane; \citealt{Djorgovski1987,Dressler1987}). 
Redshift based distance estimates are more widely available for galaxies detected by
{\it Gaia}. 
However, the accuracy of redshift based distance estimates is diminished for relatively small 
distances because the recession velocity predicted by the Hubble flow is comparable
to typical radial peculiar velocities. 
Such estimates are therefore inappropriate for very nearby galaxies ($< 10$ Mpc) that will show the
largest observer induced proper motions. 
Additionally, the Cosmicflows-3 catalog includes line-of-sight peculiar 
velocity measurements, which \cite{Graziani2019} used to derive the 
local peculiar velocity field and matter distribution. 
In Section~\ref{sec:predictions}, we employ these peculiar velocities  
to forecast the detection of secular parallax in future {\it Gaia} data
releases

We cross-matched Cosmicflows-3 with  
{\it Gaia} DR2 using a 3 arcsecond matching radius, which resulted in 9,823 objects. 
We performed cuts of the sample in order to mitigate 
contamination from Galactic stars, which will typically have larger, more significant 
proper motions than those expected for galaxies. 
We excluded any sources within 10 degrees
of the Galactic plane ($|b|\leq 10^\circ$), and any remaining sources with 
significant annual parallaxes measured by {\it Gaia} ($|p| \geq 5 \sigma_p$).
Finally, we removed galaxies within 1.5 Mpc, approximately the radius of the Local Group.
The resulting catalog contains 9,699 galaxies, of which only 429 have 
a full 5-parameter astrometric solution (measured position, parallax, and proper motion) in {\it Gaia} DR2. 
We use only these 429 galaxies with proper motion measurements below. 
In Section~\ref{sec:predictions}, we utilize the larger sample, including 
sources without proper motion measurements, to simulate end-of-mission 
proper motions and predict detection of the secular parallax dipole.  

\subsubsection{Visual Inspection}\label{sec:visual inspection}

\begin{figure*}[ht!]
    \centering
    \plotone{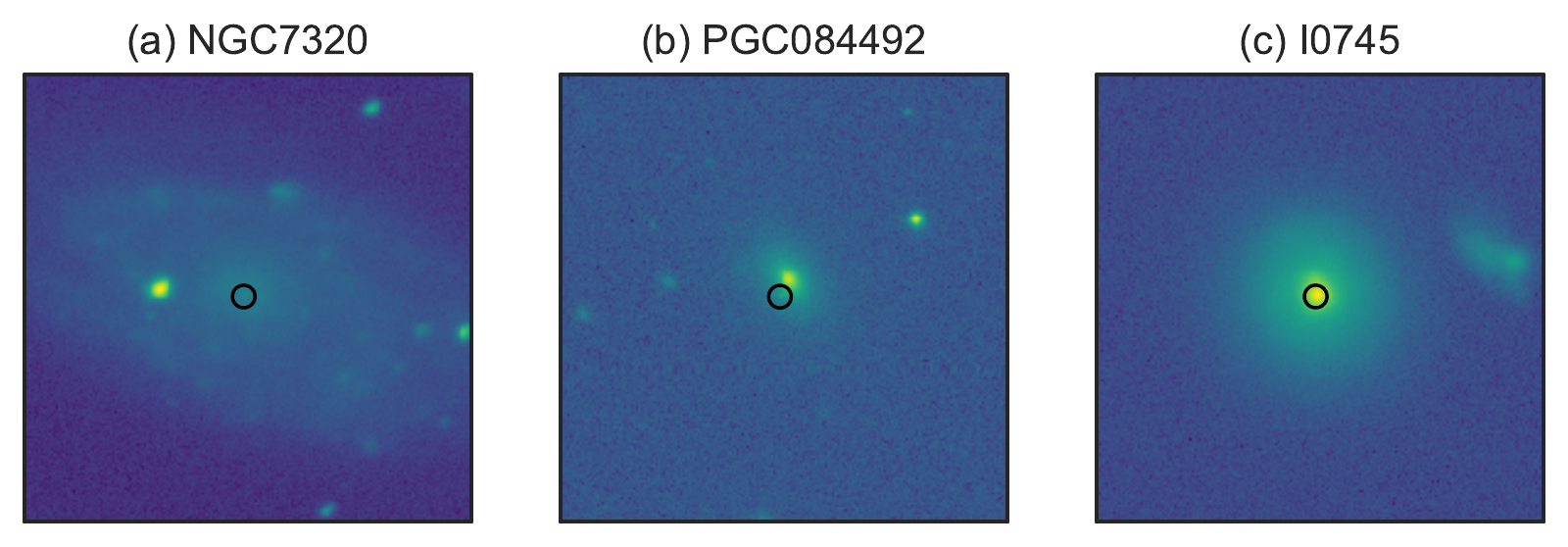}
    \caption{SDSS $g$-band images of three extended galaxies demonstrating examples 
     of the visual criteria used to verify {\it Gaia} positions. 
     Circles indicate the {\it Gaia} position for each galaxy and are each
    2 arcsec in radius (typical position errors for objects in our sample are $~0.5$ mas). (a) A galaxy with no core. (b) A galaxy where the 
    {\it Gaia} position is off-center. (c) An example of an extended galaxy with an acceptable 
    {\it Gaia} postion. } 
    \label{fig:examples}
\end{figure*}

Parallax induces the largest proper motions in nearby galaxies, but nearby galaxies are not 
point-like. However, extended objects did not receive special treatment in 
the data processing for {\it Gaia} DR2 and were treated as single stars in 
the astrometric solution \citep{Brown2018}. To ensure that the {\it Gaia} positions and proper motions 
are reasonable, we visually inspected all 429 galaxies using Sloan Digital Sky Survey (SDSS) g-band
imaging where available, or images from the Digitized Sky Survey II 
(DSS-II). 
Each image was examined for potentially problematic features, including extended galaxies
without a clear centroid or off-center {\it Gaia} positions. 
Examples of such galaxies are shown in Figure~\ref{fig:examples}.
In cases where SDSS imaging was not available and the {\it Gaia} position 
appeared marginal from DSS-II images, other images from the literature were examined. 
Nearby galaxies ($<50$ Mpc) received more scrutiny than more distant ones, since the 
proper motions of nearby objects provide a larger contribution to the secular parallax signal.
The majority of the galaxies in our initial cross-match are extended, however only
119 were identified as extended with a poor {\it Gaia} position and were 
removed from the sample. 
We identified an additional 6 {\it Gaia} positions where there were no visible sources  
as well as one foreground star, which were removed from the sample. The resulting catalog 
contains 303 galaxies which pass visual inspection.  

\subsubsection{Proper Motion and Distance Clipping}

\begin{figure}[t!]
    \centering
    \includegraphics[width=\columnwidth]{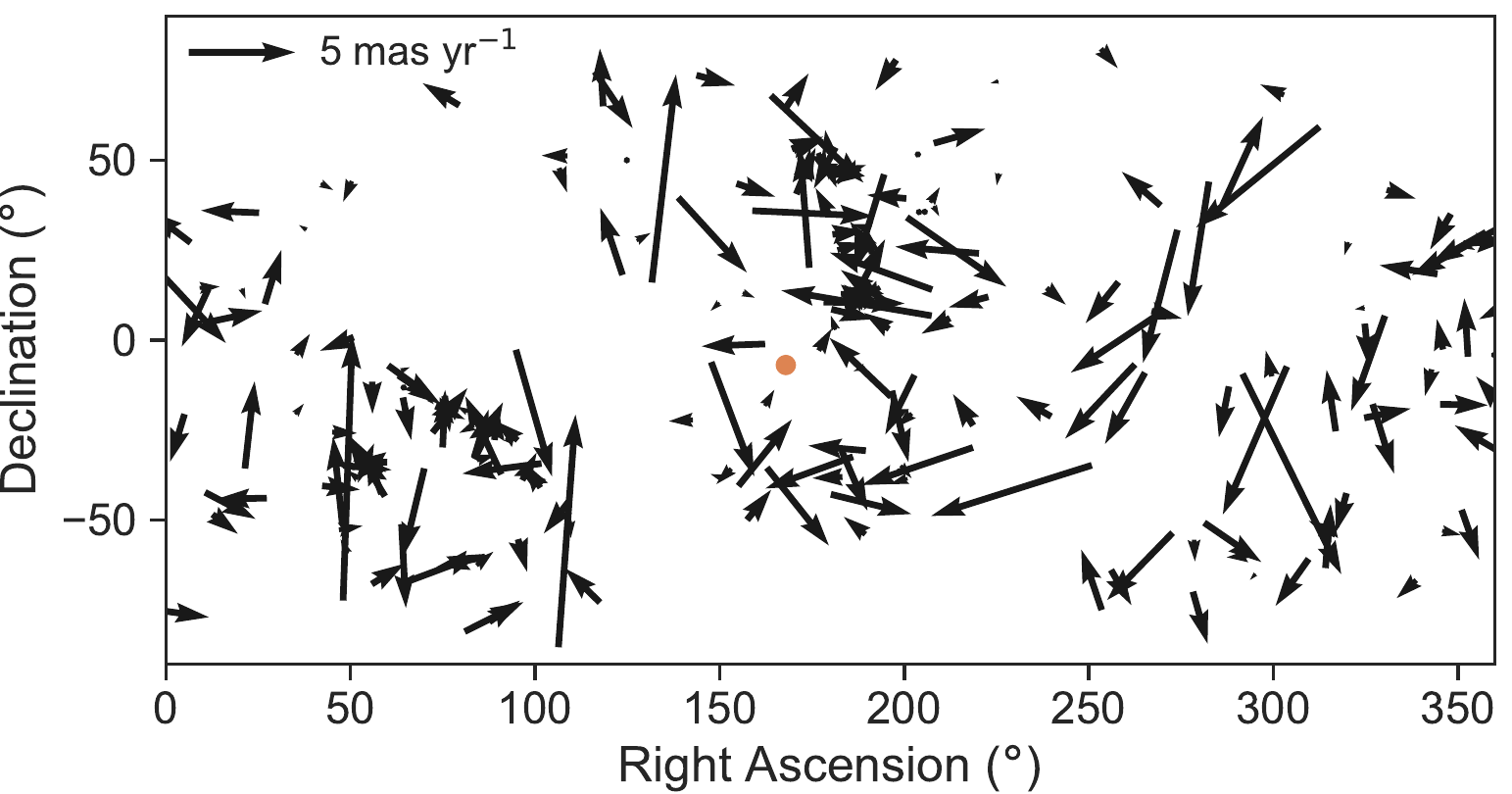}
    \caption{Proper motions and sky distribution of our sample of 232 {\it Gaia}-Cosmicflows galaxies. 
    Note that none of the proper motions shown are significant.
    The coordinates are located at the midpoint of each arrow. The direction of the 
    solar system motion with respect to the CMB is indicated by the orange dot. 
    Apparent clustering of arrows is due to the exclusion of objects within 10 degrees of 
    the Galactic plane. }
    \label{fig:PM_sky}
\end{figure}

\begin{figure}[t!]
    \centering
    \includegraphics[width=\columnwidth]{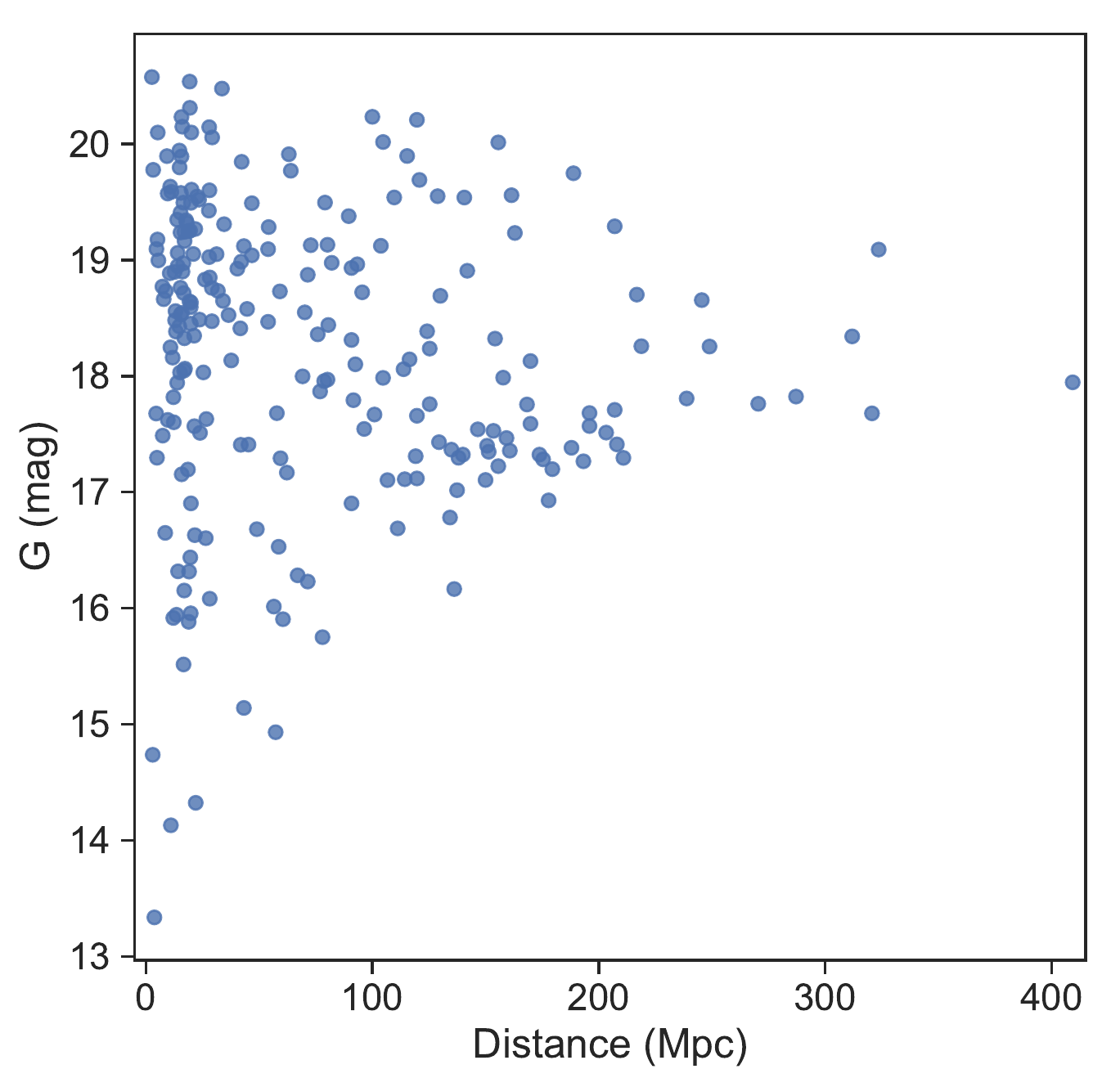}
    \caption{$G$ magnitudes vs. distance for our sample of 232 {\it Gaia}-Cosmicflows galaxies.
    {\it Gaia} proper motion uncertainties scale with magnitude.}
    \label{fig:G_dist_DR2}
\end{figure}

Out of the 303 remaining objects, 242 have proper motion 
amplitudes greater than 1 mas yr$^{-1}$, at least an order of magnitude larger than the 
expected proper motions caused by either secular parallax or a galaxy's peculiar motions. 
71 of these galaxies have $>5\sigma$ proper motions in RA and/or Dec.
As a result, an initial error-weighted fit of the parallax dipole (Equation~\ref{eqn:dipole})
to the sample has a non-significant amplitude (defined by Equation~\ref{eqn:amplitude}) 
of $32 \pm 8$ mas yr$^{-1}$ Mpc. 
Significant proper motions of galaxies are most likely either spurious measurements 
or foreground stars that were missed during visual inspection, 
so we remove any sources with $>5\sigma$ proper motions in either RA or Dec. 
The proper motions and positions of the remaining 232 galaxies are shown in 
Figure~\ref{fig:PM_sky}, and their distances and $G$ magnitudes are shown in Figure~\ref{fig:G_dist_DR2}.

\begin{figure}[t!]
	\centering
	\includegraphics[width=\columnwidth]{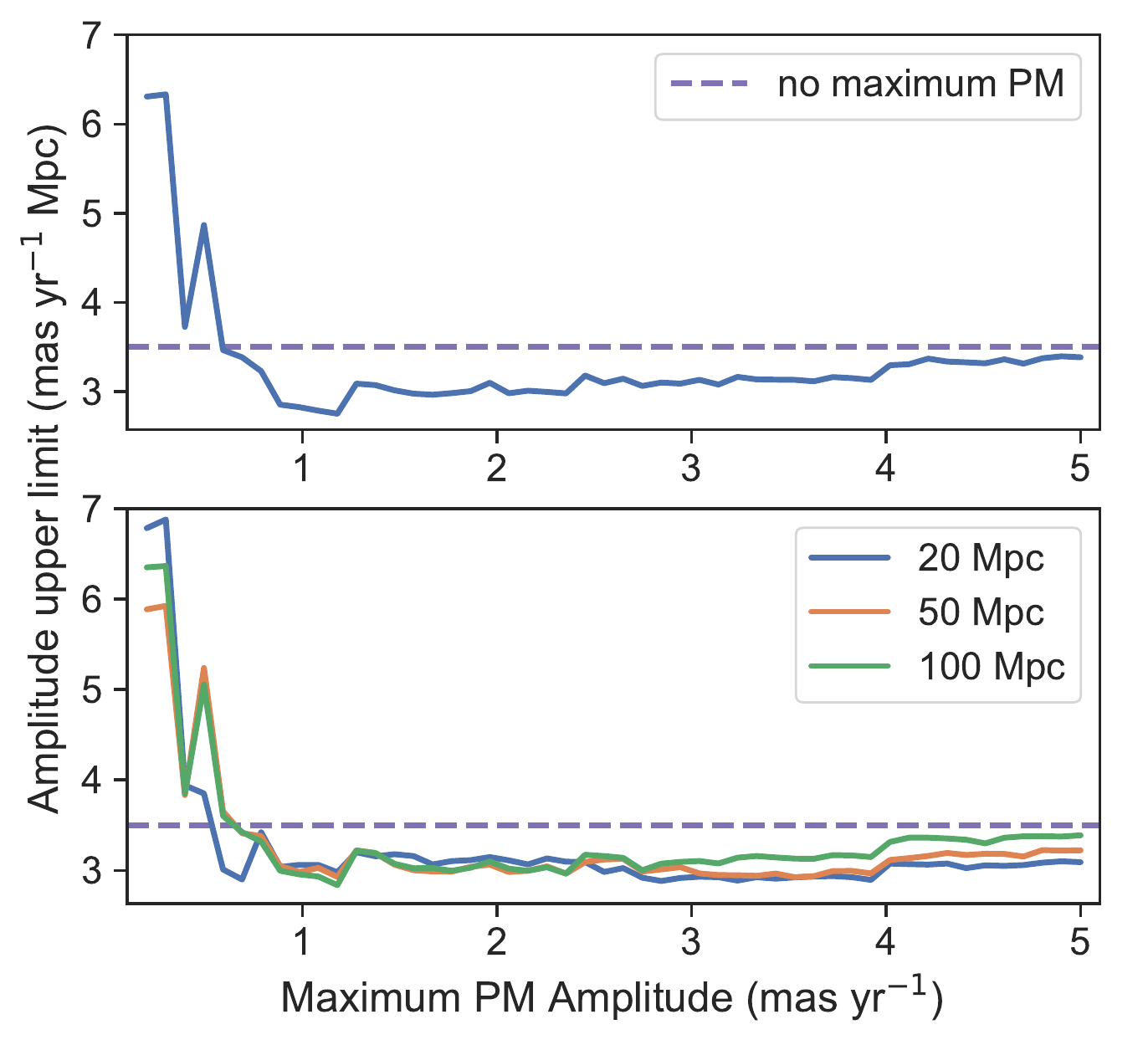}
	\caption{Top: dipole amplitude upper limits vs. maximum proper motion amplitude for the sample within 410 Mpc.
	The purple  dashed line is the upper limit for the sample
	with no maximum proper motion cut off. The sample size for the maximum cut off is 213 objects, 
	whereas the sample size for cut offs $<0.8$ mas yr$^{-1}$ is between 6 and 44 objects. 
    Bottom: Same as top for maximum distance cutoffs of 20, 50, and 100 Mpc. 
    The purple dashed line is the upper limit
    for the sample with no maximum proper motion or distance cut offs. }
	\label{fig:clipping tests} 
\end{figure}

To understand the sensitivity of our fitting technique to the remaining large, but insignificant 
proper motions, we tested
the parallax model on the sample with a range of maximum proper motion vector amplitude\footnote{
Proper motion amplitudes are calculated by 
$|\mu| = \sqrt{\mu_{\alpha*}^2 + \mu_\delta^2}$, where 
$\mu_{\alpha*} = \mu_\alpha \cos \delta$, and $\mu_\alpha$ and $\mu_\delta$ are 
the proper motion in the RA and Dec directions, respectively.}
cutoffs between 0.1 and 5 mas yr$^{-1}$. 
For each proper motion cutoff, we fit a distance dependent E-mode
dipole using least-squares minimization. 
None of the fits are significant and the coefficients are generally consistent with zero. 
For each cut we find the 95\% confidence interval upper limit for the dipole amplitude
via Monte Carlo sampling of the fit coefficients.
The top plot of Figure~\ref{fig:clipping tests} shows the amplitude upper limit
vs. maximum proper motion amplitude in the fit sample. Note that the fit amplitudes are expressed in mas yr$^{-1}$, 
so the fits are at least an order of magnitude larger than the expectation. 
The variability below the 1 mas yr$^{-1}$ cut off 
may be attributed to small sample sizes. 
The amplitude upper limit for the fit to the sample with no maximum proper motion cutoff 
is $\sim 3.5$~mas yr$^{-1}$ Mpc, and we find only a marginal reduction in the limit 
for proper motion cutoffs around 1-4 mas yr$^{-1}$.

Our galaxy sample includes objects at distances up to 410 Mpc, most of which have expected 
secular parallax proper motions smaller than a few $\mu$as yr$^{-1}$ and should not contribute
greatly to the global parallax signal relative to the more numerous nearby galaxies. 
However, the systematics that dominate the Gaia DR2 proper motions for this sample are 
distance-independent, so fits of the distance-dependent dipole may be sensitive to distance binning. 
In particular, we want to test whether the inclusion of galaxies at large distances may bias the 
best-fit amplitudes to larger values due to the dominant systematic errors. 
The bottom plot of Figure~\ref{fig:clipping tests} shows the same varying maximum proper motion
fits described above for three choices of maximum distance: 20, 50, and 100 Mpc. Again 
we find only small differences between cuts, except for low distance and proper motion cuts
where the sample size is small. Conversely, increasing the minimum distance of the sample
consistently increases the best-fit amplitudes and uncertainties. The fits are therefore most 
sensitive to the proper motions of the nearest galaxies in the sample, with little to no improvement 
on the limit from increasing the sample size by including distant galaxies. 

We next consider the impact that cluster member galaxies may have on these fits, as galaxies
in clusters can have peculiar velocities on the order of 1000 km s$^{-1}$ and may result in 
proper motions that are correlated with other members of the same cluster. 
Three nearby clusters are represented in our sample: Virgo (14 galaxies), Ursa Major (6 galaxies), 
and Fornax (12 galaxies). At approximately the distance of the Virgo cluster, a
peculiar transverse velocity of 1000 km s$^{-1}$ would correspond to a $\sim15~\mu$as yr$^{-1}$
proper motion, well below the statistical and systematic errors in the sample. If we exclude all cluster members from
the sample, we find a dipole amplitude upper limit of $\sim 3.5~\mu$as yr$^{-1}$ Mpc, 
confirming that cluster members are not the source of the large amplitudes discussed previously. 

We therefore conclude that further cuts on the catalog do not significantly improve 
the secular parallax limit. In Section~\ref{sec:limit}, we present details on this limit using the 
proper motions displayed in Figure~\ref{fig:PM_sky}.

\subsection{Results}\label{sec:limit}

\begin{figure}[t!]
    \centering
    \includegraphics[width=\columnwidth]{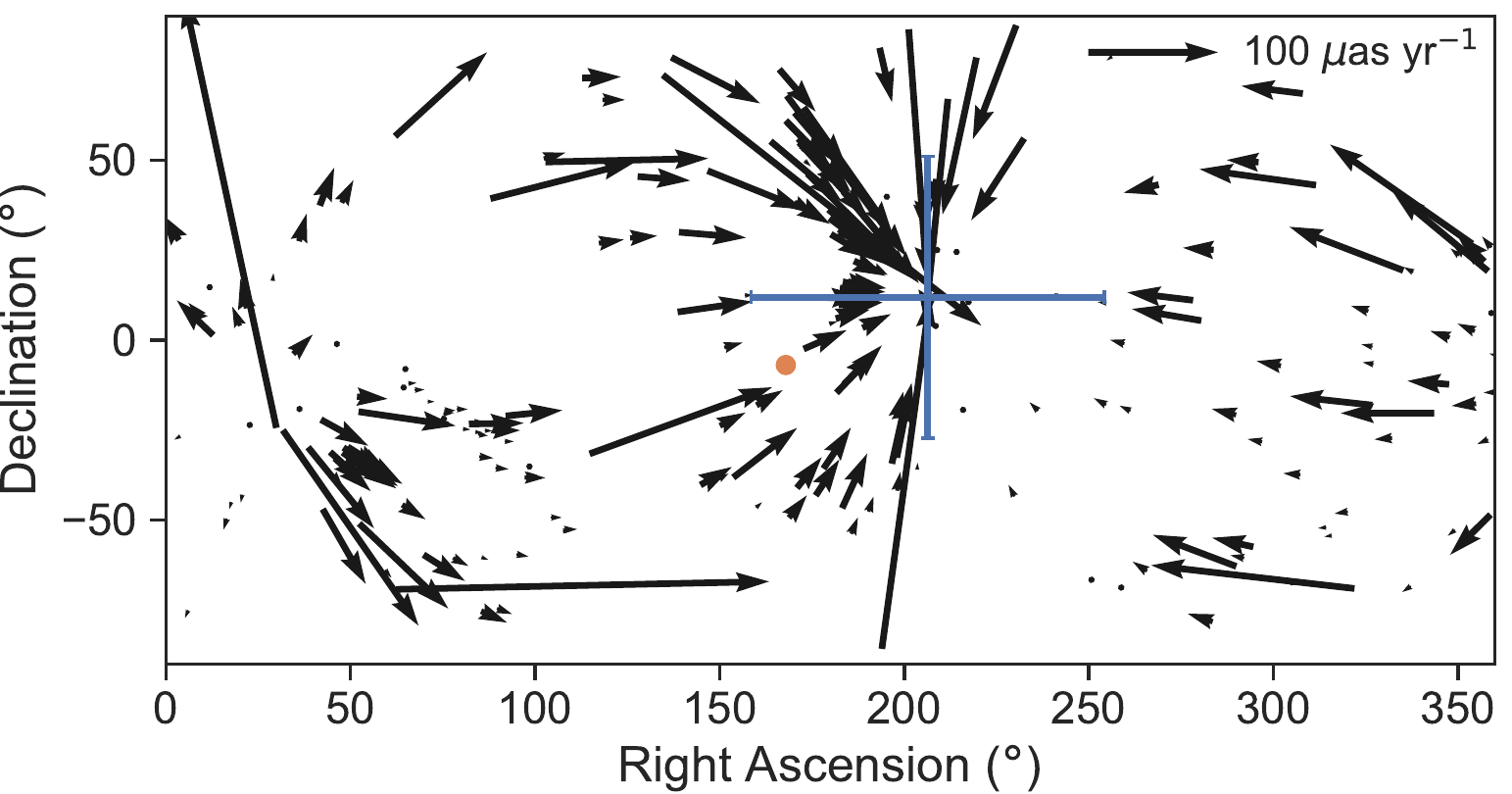}
    \caption{Best fit distance dependent dipole to the proper motions of our sample
    of 232 nearby galaxies (Figure~\ref{fig:PM_sky}). Note that this dipole is not significant.
    The blue cross shows the location and uncertainty of best fit antapex. The direction of the 
    solar system motion with respect to the CMB is indicated by the orange dot.}
    \label{fig:parallax_best_fit}
\end{figure}

We fit a distance dependent E-mode dipole to the proper motions of 
our sample of 232 galaxies using least-squares fitting. The best fit dipole is
$(s_{10}, s_{11}^{Re}, s_{11}^{Im}) = (800 \pm 3300, 2400 \pm 2100, -1200 \pm 1800)
~\mu\rm{as~yr}^{-1}$~Mpc, 
which results in a 95\% confidence interval upper limit on the dipole amplitude of $\sim3500~\mu$as~yr$^{-1}$~Mpc.
The best fit proper motions are shown in Figure~\ref{fig:parallax_best_fit}.
The dipole is not significant,
but it is coincidentally nearly anti-aligned with the expected dipole.

The limit is about an order of magnitude larger than either the expectation from the solar system's 
velocity or the peculiar velocities of nearby galaxies. It therefore represents an insignificant upper 
limit on the velocity of the solar system with respect to the bulk peculiar flow.
For an estimate of $H_0$, one needs only the component of of the proper motion caused the by 
the velocity with respect to the CMB, which may be estimated by fixing the dipole direction to the CMB 
dipole apex. The unfixed dipole is nearly anti-aligned with the CMB dipole, so the fixed-direction 
fit has zero amplitude. The uncertainties on dipole coefficients $(s_{10}, s_{11}^{Re}, s_{11}^{Im})$ are
$(270, 1500, 320)~\mu$as~yr$^{-1}$~Mpc, and the 95\% confidence upper limit on the amplitude
is $\sim 1500~\mu$as yr$^{-1}$ Mpc. The fractional uncertainty of the amplitude may be translated to 
$H_0$, which implies an $H_0$ limit of 1400 km s$^{-1}$ Mpc$^{-1}$. We note that we do not measure 
an $H_0$ limit; rather it is an estimate of the limit that may be achieved using current {\it Gaia} data.

The large scale systematics in {\it Gaia} DR2 proper motions are of the
order 40 $\mu$as yr$^{-1}$  for angular scales $\gtrsim 18^\circ$ \citep{Lindegren2018, Mignard2018}. 
Further, \cite{Mignard2018} find that the dipole systematic proper motion effects 
are of the order of 10 $\mu$as yr$^{-1}$ for their {\it Gaia} celestial 
reference frame quasar sample. 
To assess the large scale systematics present in our sample of 232 galaxies, 
we fit a distance-independent E1 dipole to the sample. 
We find an insignificant dipole 
of $(s_{10}, s_{11}^{Re}, s_{11}^{Im}) = (280 \pm 210, 360\pm170, -80\pm160)
~\mu\rm{as~yr}^{-1}$, which corresponds to $200\pm80~\mu$as yr$^{-1}$ amplitude.
The apex directions of the distance-dependent and independent fits are consistent, indicating
that both fits probe the same underlying systematics. 
Additionally, the distance-dependent dipole amplitude normalized to the median distance of the 
sample, 43 Mpc, is $\sim 30~\mu$as yr$^{-1}$, which is consistent with the known {\it Gaia} systematics. 

Either an improvement of the statistical and systematic errors or a larger sample size will therefore be 
required to reduce the uncertainty of the secular parallax limit. 
The uncertainty on the coefficients and amplitude of the global signal scales with the sample size as $N^{-1/2}$,
so to achieve uncertainty on the signal at 1 Mpc of about 10 $\mu$as yr$^{-1}$ 
without any reduction on the individual proper motion uncertainties would
require a larger sample size by a factor of 10,000, assuming identical distance 
and sky distributions. However, proper motion uncertainties scale as $t^{-3/2}$, 
so the longer time baseline of future {\it Gaia} data releases will decrease 
the individual uncertainties. 
The average per-object uncertainties will decrease by about an order of magnitude by 
{\it Gaia}'s nominal end of mission.
For the sample used to make the above limit, the mean expected uncertainties are 
$\sim 120$ and $110~\mu$as yr$^{-1}$ in RA and Dec.

\section{Secular Parallax Predictions}\label{sec:predictions}

As demonstrated in the previous section, 
the systematics present in {\it Gaia} DR2 act as a noise floor below which we may not analyze low 
multipole proper motions. Future {\it Gaia} data releases, however, will likely contain 
larger numbers of extragalacitic proper motion measurements with lower
per object statistical uncertainties and systematics. 
We can expect that the majority of the 9,823 objects in our initial 
{\it Gaia}-Cosmicflows crossmatch will have proper motion measurements
by the final data release. 
In this section, we therefore forecast 
the detectability of secular parallax for {\it Gaia} end-of-mission astrometry
and explore the ideal sample selection to be used for future data releases. 

\subsection{Simulated Proper Motion Catalog}\label{sec:sims_catalog}

We calculate proper motions consistent with {\it Gaia's} expected 
end-of-mission performance for objects in the  {\it Gaia}-Cosmicflows crossmatch described in Section~\ref{sec:dr2_selection}.  We exclude any galaxies closer than
1.5 Mpc, with galactic latitude below 10 degrees, or with significant annual
parallax detected by {\it Gaia}. The later two requirements mitigate contamination
by foreground stars. One additional object was identified as a foreground star 
by visual inspection. The catalog contains 9,698 objects, most of which do not have 
measured proper motions as of {\it Gaia} DR2.
For each galaxy, we simulate a {\it Gaia} end-of-mission proper motion 
based on three components: the source's predicted proper motion errors, the expected secular parallax, 
and predicted peculiar velocity based on the Cosmicflows-3 peculiar velocity field. 

End-of-mission proper motion errors are calculated using the PyGaia Python 
toolkit.\footnote{\url{www.cosmos.esa.int/web/gaia/science-performance}}
The predicted uncertainties assume a five year mission, although {\it Gaia}'s 
total mission lifetime has already passed its nominal five years \citep{GaiaCollaboration2016}.
The calculation depends on the source's $G$ magnitude, ecliptic latitude, and 
$V-I_C$ color. We set the latter to zero for all sources as the color has negligible
impact on the predicted uncertainty and is not available for all sources. 
The mean expected uncertainties are 77 $\mu$as yr$^{-1}$ for proper motion 
in R.A. and 68 $\mu$as yr$^{-1}$  in Dec. 
Note that the predicted uncertainties do not include potential systematic errors.  
We generate proper motion noise for each object by randomly sampling 
from a Gaussian distribution with standard deviation set to the object's
predicted proper motion uncertainty in R.A. and Dec.

Peculiar proper motions are calculated from the local peculiar velocity field
described in detail in \citet{Graziani2019}. The velocity field was reconstructed from 
the Cosmicflows-3 catalog, utilizing the observed distance moduli and redshifts 
of Cosmicflows galaxies to infer both the matter over-density field and the three-dimensional
peculiar velocity field. We find peculiar proper motions then by converting the transverse
velocities to angular motions. The transverse velocities are on the order of a few hundred km s$^{-1}$ in the CMB frame, 
with a maximum of $\sim 1000$ km s$^{-1}$. The resulting peculiar proper motions
range from $0.005$ to $60~\mu$as yr$^{-1}$ and the mean is $\sim 1~\mu$as yr$^{-1}$. 

\begin{figure}[t!]
    \centering
    \includegraphics[width=\columnwidth]{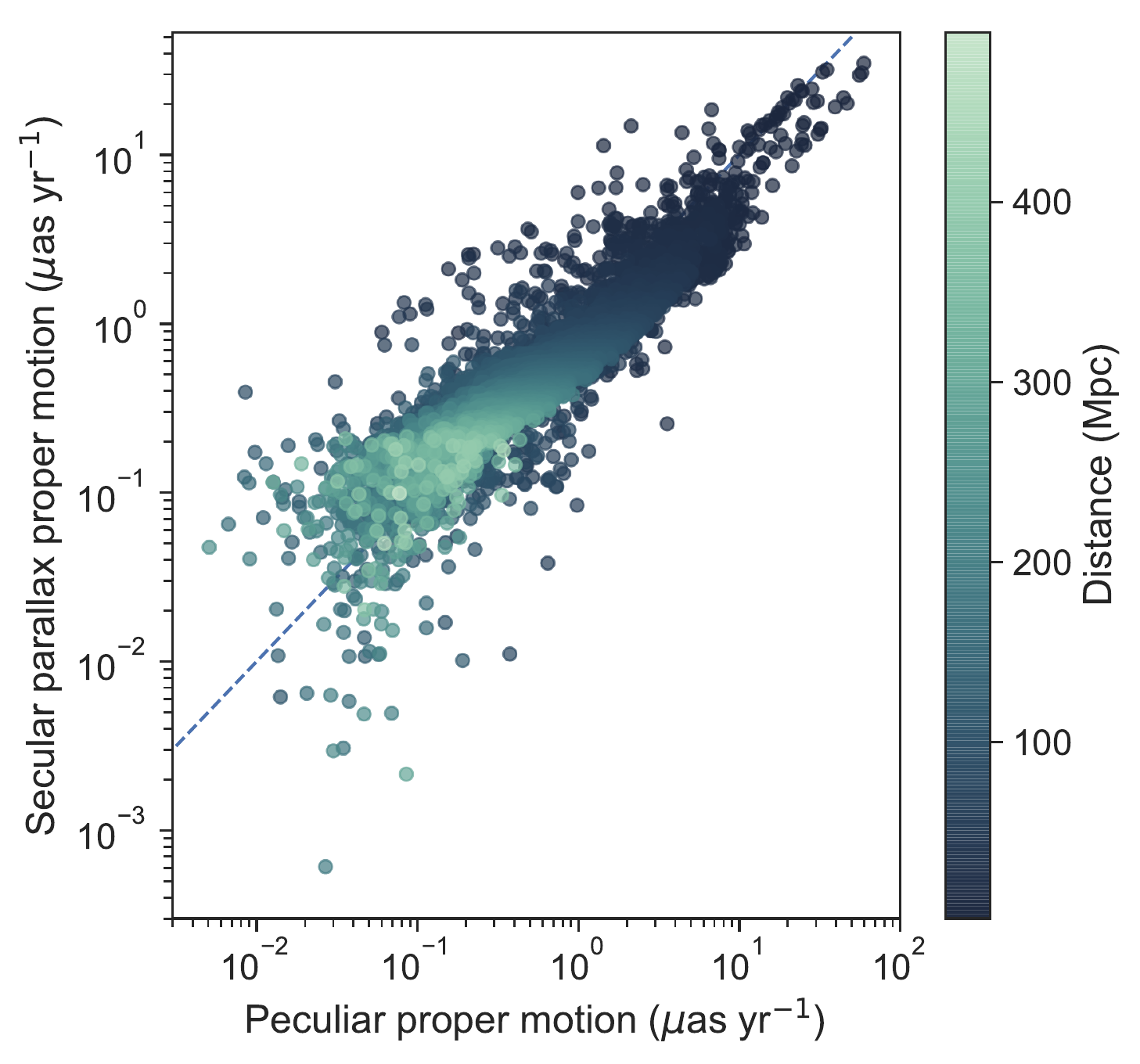}
    \caption{Simulated secular parallax vs. peculiar proper motion components for individual 
    objects. Colors scale with the distance of each galaxy. 
    For reference, the dashed line denotes a one-to-one correspondence 
    where peculiar proper motions show the same amplitude (but not necessarily direction) 
    as the secular parallax.}
    \label{fig:pm_components}
\end{figure}

In Figure~\ref{fig:pm_components}, we plot the peculiar vs. 
secular parallax proper motions (calculated from Equation~\ref{eqn:dipole}) for each galaxy. 
The peculiar proper motion
amplitude is larger than the expected secular parallax for 58\% of galaxies in the sample. 
While the transverse velocity amplitudes of individual galaxies do not depend on 
distance, the corresponding peculiar angular motions depend on distance as $1/D$, which 
complicates the measurement of secular parallax.
Additionally, the transverse velocity angular power spectrum varies with distance, with more 
power in $\ell =1$ at smaller distances (\citealt{Hall2018}; Section~\ref{sec:LSS}). 

To create the simulated proper motion catalog, we first sample each object's end-of-mission 
uncertainties to generate a noise term, and then add the predicted parallax and peculiar proper
motions. We note that the largest contribution to real extragalactic proper motions is 
typically the secular aberration drift, which we do not include in our simulations in this work. 
However, the secular aberration drift is expected to be measured with high significance
using the sample of $>5\times10^5$ AGN detected by {\it Gaia} \citep{Paine2018}. 
The following predictions are therefore made with the assumption that the secular
aberration drift dipole can be constrained or subtracted prior to fitting for secular parallax. 
We also do not include the effects of possible contaminating foreground stars; 
the proper motion is calculated assuming that each {\it Gaia} source is the Cosmicflows galaxy. 

\subsection{Gaia End of Mission Simulations}\label{sec:sims}

\begin{figure}[t!]
    \centering
    \includegraphics[width=\columnwidth]{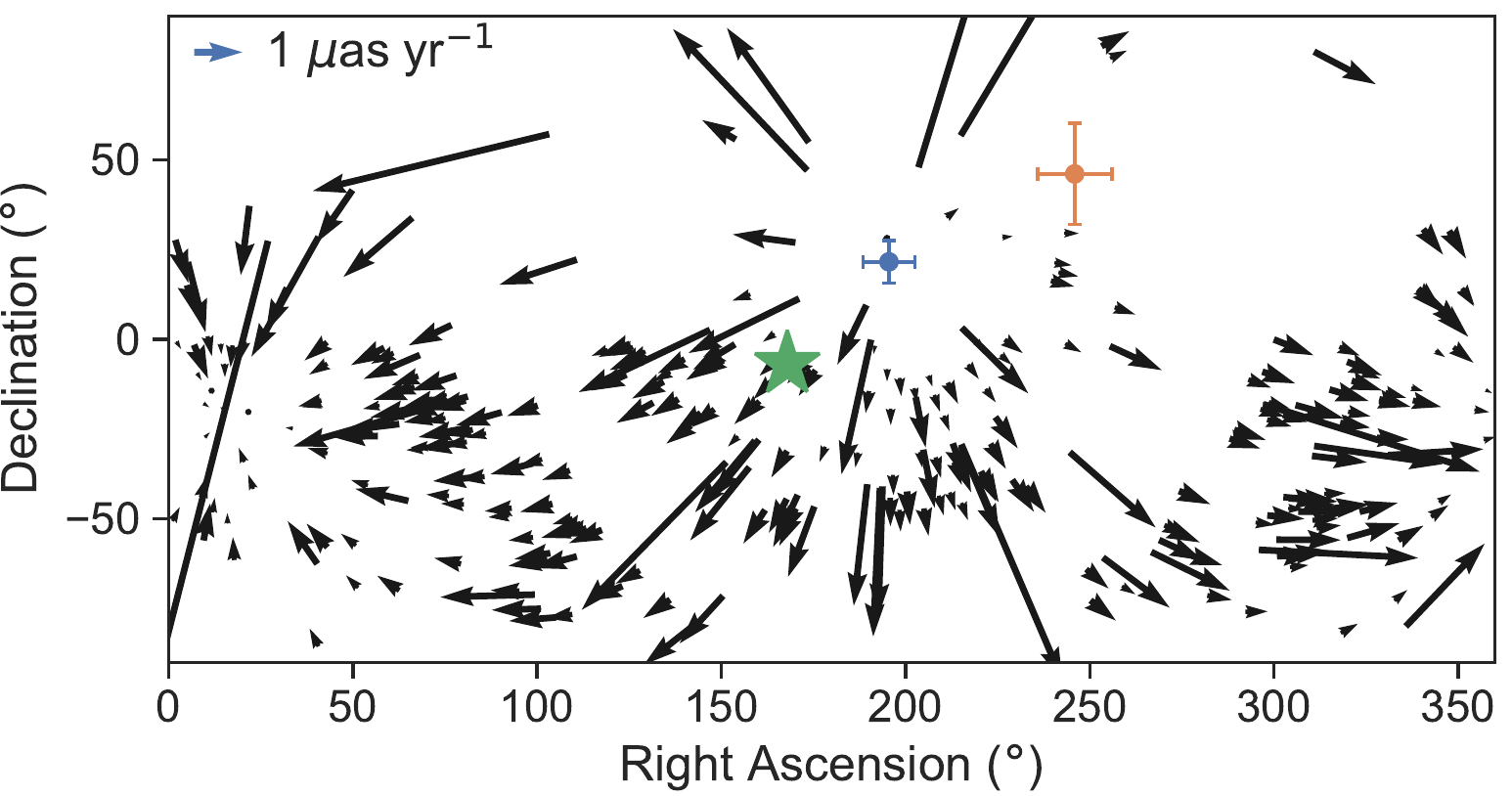}
    \caption{Arrows show the simulated best fit dipole for the proper motions including 
    secular parallax, peculiar motions, and noise consistent with {\it Gaia} end-of-mission uncertainties. 
    The midpoint of the arrows represent the coordinates
    of 300 objects randomly selected from the simulated catalog to illustrate the proper 
    motion vector field. The blue cross shows the 
    location of the apex of the fit, whereas the orange cross shows the location
    of the apex when fitting to the peculiar proper motions alone, and the green
    star is the location of the CMB apex and the apex of the secular parallax dipole.
    The offset of the best fit dipole from the CMB apex demonstrates the mixing 
    of the peculiar and parallax dipoles.}
    \label{fig:unconstrained_fit}
\end{figure}

To simulate secular parallax detection, we perform error weighted least squares 
fits of the distance-dependent dipole model to the simulated proper motion catalog. 
We perform 1,000 fits, where for each trial we generate random noise for each object 
by sampling the end-of-mission uncertainties. The average resulting fit is a 
$107 \pm 12~\mu$as yr$^{-1}$ Mpc dipole with apex of 
($195\pm6^\circ$, $22\pm7^\circ$) in RA and Dec, 
and mean Z-score of 7.8. The mean dipole coefficients are listed in Table~\ref{tab:sim_fits}.
The average best-fit dipole is significantly offset from the secular parallax expectation from the CMB, 
which is due to dipole correlations of the peculiar proper motions in our sample. 
For each trial, we also perform fits to the catalog excluding the predicted secular parallax
proper motions, in order to demonstrate the impact of peculiar motions on secular parallax detection. 
The mean dipole coefficients for the noisy peculiar proper motions are also listed in Table~\ref{tab:sim_fits}
and correspond to a $69.1 \pm 11.6 ~\mu$as yr$^{-1}$ Mpc dipole
toward ($246\pm14^\circ$, $46\pm10^\circ$). The dipole detected in the full simulation
reflects the combination of the expected secular parallax and peculiar proper motions. 
In fact, the best-fit dipole parameters are consistent with the sum of the separate parallax and 
peculiar dipole parameters, and the amplitude is consistent with the parallax and peculiar dipole
amplitudes summed in quadrature. The mixing of the two dipoles is further demonstrated in 
Figure~\ref{fig:unconstrained_fit}, which shows the locations of the best-fit dipole 
apex for the full simulation, for the peculiar proper motions, and the expected 
secular parallax apex (CMB apex).

\begin{deluxetable*}{rccccc}

\tablehead{\nocolhead{} & \colhead{ $s_{10}$} & \colhead{$s_{11}^{Re}$} & \colhead{$s_{11}^{Im}$} & \colhead{Amplitude} & \colhead{Apex coordinates} \\
\nocolhead{} & \colhead{($\mu$as yr$^{-1}$ Mpc)} & \colhead{($\mu$as yr$^{-1}$ Mpc)} & \colhead{($\mu$as yr$^{-1}$ Mpc)}& \colhead{($\mu$as yr$^{-1}$ Mpc)} & \colhead{RA, Dec}}

\startdata
 Secular Parallax & 27.2 & $-155$ &  $-33.1$ & 78 & $167.9^\circ$, $-6.9^\circ$\\
 Peculiar & $-139(38)$ &  $-38.9(24.0)$ & $86.9(20.3)$ & $69.1(11.6)$ & $245.9^\circ$, $46.1^\circ$\\
 Full simulation & $-112(38)$ &  $-194(24)$ & $53.8(20.3)$& $107(12)$ & $195.5^\circ$, $21.6^\circ$
 \enddata
 \caption{Dipole properties for the expected secular parallax dipole, the best-fit 
 dipole to simulated peculiar proper motions, and the best-fit dipole to the full simulation 
 including secular parallax, peculiar, and random noise proper motion components.
 }
 \label{tab:sim_fits}
\end{deluxetable*}

From Equation~\ref{eqn:dipole}, one can demonstrate that 
the vector field produced by addition of two dipoles is simply another dipole. 
It is therefore not possible to separate the parallax and peculiar dipoles 
observationally without prior knowledge of either component. 
However, the CMB dipole has a well measured direction, so we may 
fix the location of the dipole apex in order to recover the secular parallax amplitude. 
This is achieved by constraining the ratios of $s_{10}$ to $s_{11}^{Re}$ and $s_{11}^{Im}$
while fitting (the signs and relative absolute values of the coefficients determine the direction 
of the dipole, whereas the sum of squares gives the amplitude as in Equations~\ref{eqn:power}-\ref{eqn:amplitude}). We perform 1,000 trials fitting the fixed-direction distance-dependent
dipole, again randomly generating noise for each trial. The resulting average 
fit has amplitude $\sim$ 74 $\mu$as yr$^{-1}$ Mpc detected
with mean Z-score of 9.5. This represents a significant detection of the secular 
parallax amplitude. 
We note that the peculiar proper motion dipole is nearly orthogonal to the expected
secular parallax dipole and therefore contributes almost zero amplitude to the fixed-direction 
fits. The orthogonality of the two components is a fortunate coincidence due to the 
solar system's phase in its Galactic orbit and the Galaxy's peculiar motion.  

Finally, we test the possibility of constraining both the parallax and peculiar dipoles.
We perform 1,000 trials simultaneously fitting two distance-dependent dipoles, one with
direction fixed to the CMB apex and one with variable direction and amplitude. 
Neither dipole is recovered when simultaneously fitting, so it will be necessary to account 
for the secular parallax dipole prior to studying the peculiar proper motions from LSS. 
This may be achieved by either fitting for the parallax dipole independently as described above 
and subtracting the detected field, or assuming the secular parallax proper motion field 
from the CMB.

\subsubsection{Predicted Dependence on Sample Selection}

The 9 $\sigma$ detection prediction is made for nearly the full {\it Gaia}-Cosmicflows sample, 
only employing cuts on the sample to mitigate stellar contamination and poor centroid fits. 
However, the predicted detection is very sensitive sample selection. Below, we consider 
several variables that may impact the detection: the minimum and maximum distances, 
$D_{min}$ and $D_{max}$; maximum individual proper motion amplitude;  and minimum
$G$ magnitude. 

\begin{figure}[t!]
    \centering
    \includegraphics[width=\columnwidth]{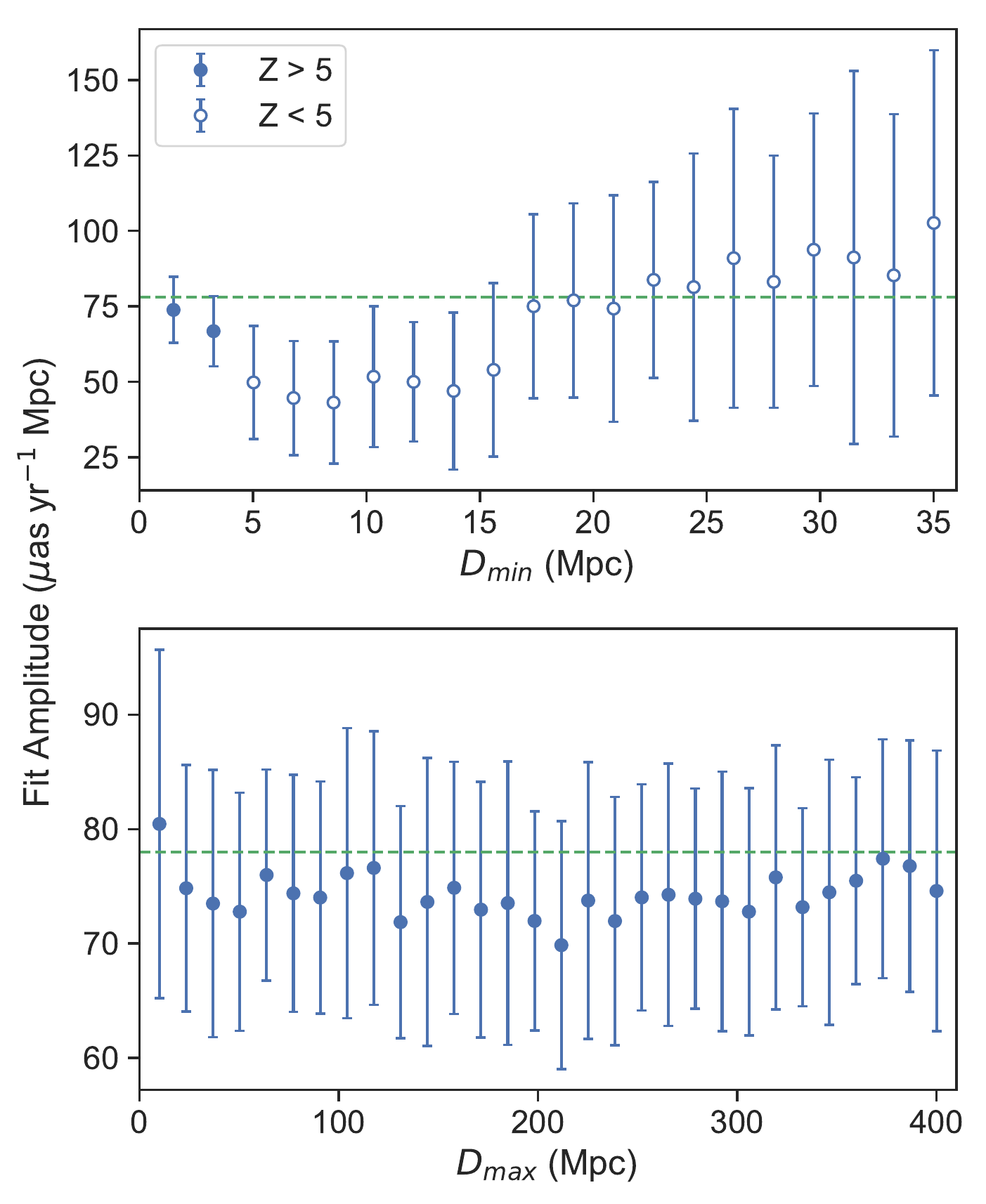}
    \caption{Simulated fixed-direction, distance-dependent dipole fits  
    for minimum distance cuts between 1.5 and 35 Mpc (top) and for maximum
    distance cuts between 10 and 400 Mpc (bottom). 
    Fits are made to the simulated proper motions including expected 
    secular parallax, model predicted peculiar motions, and noise components. 
    The green dashed lines indicate the input secular parallax amplitude of 78 $\mu$as yr$^{-1}$. 
    Top: the first point indicates a $\sim 10\sigma$ detection of the 
    secular parallax amplitude. 
    Note that only the first two distance cuts result in significant fits. 
    Bottom: all fits are significant regardless of maximum distance cut.
    The uncertainties are similar to the first two points in the top plot (note that
    the axes scales are not the same). }
    \label{fig:r_cuts}
\end{figure}

The choice of $D_{min}$ should impact the relative contribution of the peculiar proper motions
to the detected dipole since the peculiar velocity dipole decreases in power for larger distances
(\citealt{Hall2018}, see also Section~\ref{sec:LSS}). 
The peculiar dipole is approximately orthogonal to the expected secular parallax 
in all distance bins, and should therefore have nearly zero component in the fixed-direction fits.
In Figure~\ref{fig:r_cuts}, we show the fixed-direction best-fit
dipole amplitudes for $D_{min}$ cuts between 1.5 and 35 Mpc, where we performed 50 trials 
per distance cut. Only $D_{min}<5$ Mpc cuts result in significant detection of the secular parallax amplitude 
on average. The mean Z-scores of the fits for the two smallest $D_{min}$ cuts are 9.5 and 8.0, respectively. 
For $D_{min}$ cuts between 20 and 35 Mpc, typical Z-scores are $~2$. 
The prediction is therefore very sensitive to the inclusion of very nearby galaxies
which have predicted parallaxes between 8 - 35 $\mu$as yr$^{-1}$.  
By comparison, $D_{max}$ has little impact on the detection. 
Figure~\ref{fig:r_cuts} also shows the best-fit dipole amplitudes for $D_{max}$ cuts
between 10 and 400 Mpc.  We find that the best-fit 
dipole amplitude and Z-score are fairly constant for all $D_{max}$. 
The significance of the fits drops to $\sim7\sigma$ for distance cuts $<20$ Mpc, 
where the sample size is $<500$ objects.  
Distant galaxies with large uncertainties compared to their expected 
parallax therefore have little impact on the detection as long as nearby objects with the 
most signal are included. 

\begin{figure}[t!]
    \centering
    \includegraphics[width=\columnwidth]{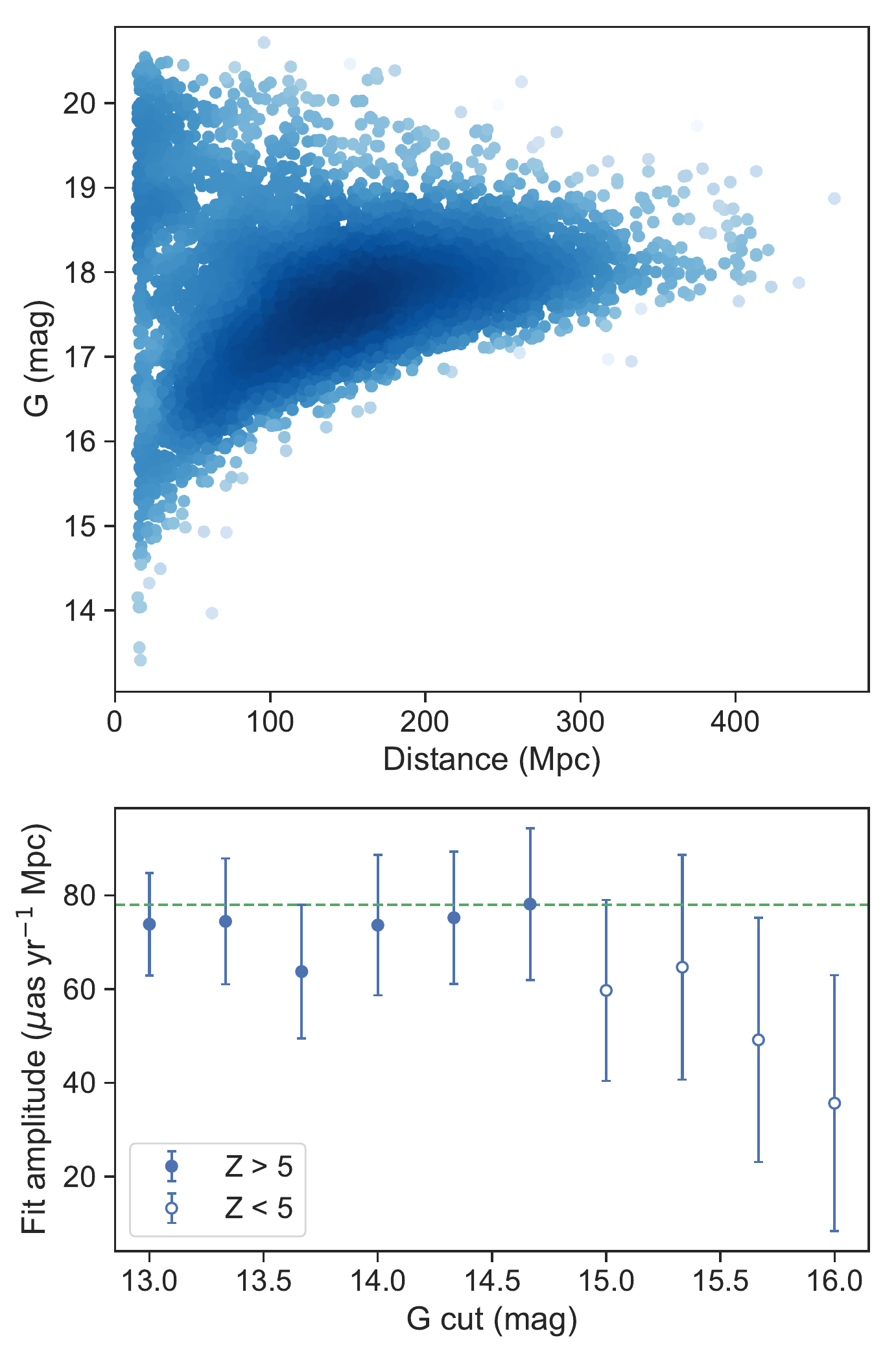}
    \caption{Top: $G$ magnitudes vs. distance for {\it Gaia}-Cosmicflows galaxies.
   	Predicted {\it Gaia} end-of-mission proper motion uncertainties scale with $G$.
    Bottom: simulated fixed-direction, distance-dependent dipole fits  
    for minimum $G$ magnitude cuts between 13 and 16 mag.  
    The green dashed line indicates the input secular parallax amplitude of 78 $\mu$as yr$^{-1}$. 
    The first 6 cuts (minimum $G<15$) result in significant average fits.  }
    \label{fig:G_cut}
\end{figure}

The $G$ magnitude of each object impacts the simulated proper motions 
due to the dependence of the end-of-mission proper motion uncertainties on $G$. 
Magnitude vs. distance
for each object are plotted in Figure~\ref{fig:G_cut}, showing a tail of nearby, bright galaxies
with $G<15$. These objects will have the largest predicted proper motions and smallest 
uncertainties, provided that good astrometric solutions can be obtained for nearby, extended galaxies. 
The error-weighted fits are therefore highly sensitive to proper motions of these galaxies. We test this by varying the minimum 
$G$ magnitude in the sample between 13 and 16 mag, and for each cut we perform 50 fits
with resampled noise components.
The best-fit dipole amplitudes are shown in Figure~\ref{fig:G_cut}. We find that all fits
including galaxies brighter than 15 mag are significant and consistent with secular parallax 
within 1 $\sigma$. 
Relatively bright and nearby galaxies will most likely require closer individual inspection when 
real {\it Gaia} data is available in order to cull objects that may throw off the total fits 
due to spurious proper motions or larger than expected peculiar motions. 
The the fact that some fits for $G$ cuts $<15$ are significant indicates that secular 
parallax may be detectable only if the majority of bright galaxies need not be culled. 
Nearby, diffuse galaxies will also require individual inspection, but they do not contribute 
to the measured signal as much as the brightest galaxies.

\begin{figure}[t!]
    \centering
    \includegraphics[width=\columnwidth]{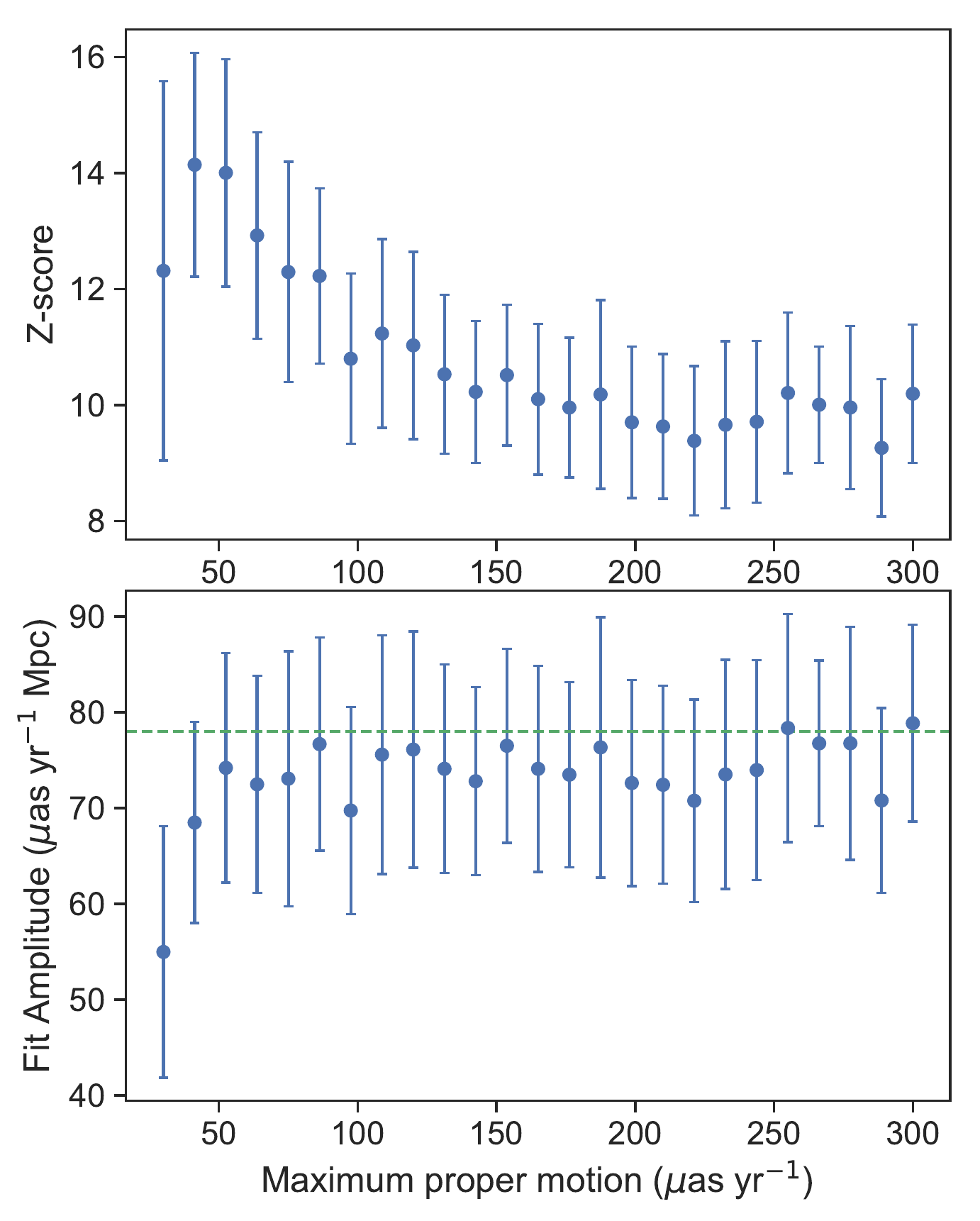}
    \caption{Z-scores (top) and best-fit amplitudes (bottom) of fixed-direction, distance-dependent
    dipole fits for maximum proper motion cuts up to 300 $\mu$as yr$^{-1}$. 
    The green dashed line indicates the input secular parallax amplitude of 78 $\mu$as yr$^{-1}$. }
    \label{fig:max_pm}
\end{figure}

The mean noise-free proper motion is $\sim 1~\mu$as yr$^{-1}$ and the maximum is 
65 $\mu$as yr$^{-1}$, whereas the mean expected uncertainty is about 80 $\mu$as yr$^{-1}$.
The majority of simulated proper motions are then dominated by noise 
and vary greatly between trials. Controlling the maximum individual proper motion
can therefore probe the dependence of the fits on noisy, insignificant proper motions.  
In Figure~\ref{fig:max_pm}, we plot the best-fit fixed-direction dipole amplitudes 
and Z-scores for 
maximum individual proper motion amplitude cuts up to 300 $\mu$as yr$^{-1}$, 
where we have performed 50 trials per proper motion cut. 
New noise components are randomly generated for each trial, so the specific objects that are cut
vary per trial. 
The best-fit dipole amplitude only depends on proper motion cut if the maximum 
proper motion is small enough so as to exclude nearby galaxies with the largest secular parallaxes. 
However, the significance of the fits is increased for cuts between 50-150 $\mu$as yr$^{-1}$.
Picking a proper motion cut in this range does not impact 
the previously established dependence on distance and magnitude cuts: the fits are still only
significant if relatively bright galaxies within 5 Mpc are included, but are insensitive to $D_{max}$.

\section{Low Multipole Analysis of Peculiar Proper Motions}\label{sec:peculiar}

In this section, we study the predicted peculiar proper motions of Cosmicflows galaxies and 
compare to LSS theory. We also assess the impact that peculiar proper motions may 
have on quadrupole ($\ell=2$) proper motion measurements, including gravitational waves and 
anisotropic expansion. In the following subsections, we utilize the simulated 
catalog described in Section~\ref{sec:sims_catalog} including only
the peculiar velocity component of each galaxy's proper motion. 

\subsection{Comparison to LSS Theory}\label{sec:LSS}

Peculiar velocities arise from gravitational interactions with large-scale matter density fluctuations, 
causing transverse velocities, and therefore proper motions, of galaxies to deviate from a perfectly 
isotropic Hubble flow. Previous works (e.g. \citealt{Darling&Truebenbach2018} and \citealt{Hall2018})
have demonstrated that extragalactic proper motions may probe these peculiar transverse velocities 
and therefore the matter power spectrum. 
\cite{Hall2018} derive the peculiar velocity power spectrum for low redshift based on the matter power
spectrum computed with the  CAMB cosmology code. 
The peculiar velocity field is expected to be nearly curl-free and corresponds to an E-mode 
transverse velocity power spectrum where the dominant mode is distance dependent.  
Dipole correlations dominate for small distances, and the power transfers to higher $\ell$ (smaller angular scales) 
with increasing distance. 

\begin{figure}[t!]
    \centering
    \includegraphics[width=\columnwidth]{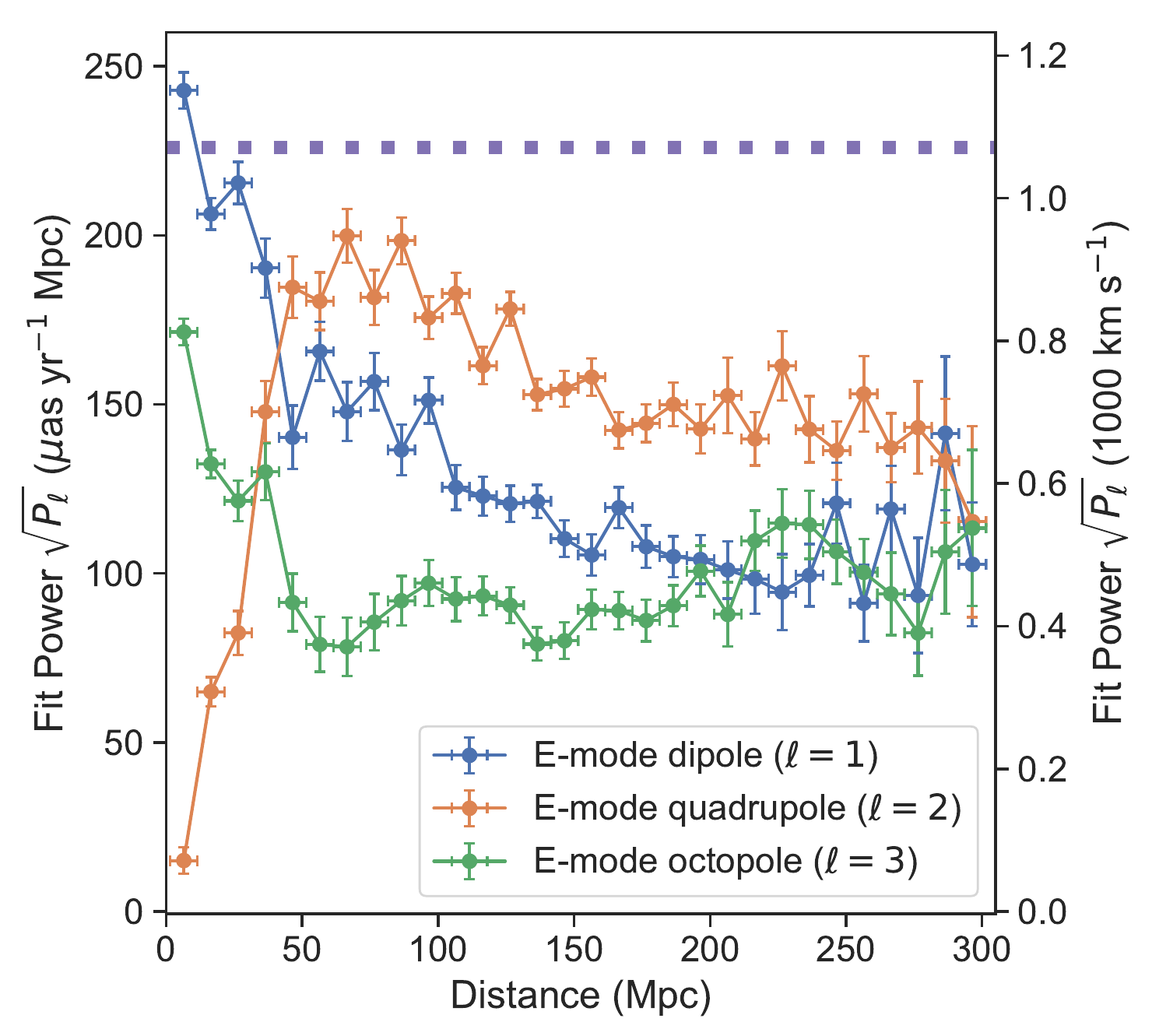}
    \caption{E-mode distance-dependent dipoles (blue), quadrupoles (orange), and 
    octopoles (green) fit to the peculiar proper motions for objects in 10 Mpc distance bins
    up to 300 Mpc. The secular parallax power is indicated by the purple dashed line.
    The right vertical axis shows the same powers converted to distance independent velocities. 
    Error bars represent the width of distance bins (horizonal) and the uncertainties of the fits (vertical), 
    The fits were performed for noise-free modeled proper motions, so the fits and uncertainties 
    are not indicative of expected $\it Gaia$ end-of-mission results.}
    \label{fig:vpec_E1E2E3}
\end{figure}

For comparison, in Figure~\ref{fig:vpec_E1E2E3} we show the E-mode dipole, quadrupole, and octopole 
square root powers ($\sqrt{P_\ell}$) fit to the noise-free model-predicted peculiar proper motions for 10 Mpc distance bins up 
to 300 Mpc, which gives an estimate of the power as a function of distance. 
The proper motion powers are normalized to 
1 Mpc and can therefore be related to a transverse peculiar velocity, where a 1 $\mu$as yr$^{-1}$ 
proper motion at 1 Mpc is equivalent to a 1 AU yr$^{-1} \approx 4.74$ km s$^{-1}$ velocity. 
This method is equivalent to fitting distance-independent 	vector fields to the transverse velocities. 
The total  velocity power in $\ell =1$, 2, and 3 is on the order of 1000 km s$^{-1}$ at all distances. 
Square root powers should not be interpreted as representative proper motions or velocities. The amplitudes 
of the fits, however, represent the maximum magnitude of the vector field.  For a dipole, the amplitude is simply 
proportional to the power, as given by Equation~\ref{eqn:amplitude}, but the amplitude for higher $\ell$ is not. 
We find the amplitude of each $\ell = 2,3$ fit by solving numerically for the maximum of the best fit vector
field magnitudes. The amplitudes of the highest power bins are 84, 88, and 80 $\mu$as yr$^{-1}$ Mpc, 
corresponding to velocities of 398, 417, and 380 km s$^{-1}$ for $\ell=1$, 2, and 3, respectively, 
roughly as expected for peculiar velocities induced by LSS. 

As expected, the dipole power decreases with distance 
as the power shifts to the quadrupole, which peaks between 50-100 Mpc. 
The octopole power, however, does not match the theory predictions, since we see more 
power in the octopole than the quadrupole at small distances. This discrepancy may 
be due to uneven sky coverage of the galaxy sample. Alternatively, the octopole power
may reflect real deviations of local matter density fluctuations, and therefore the peculiar 
velocity field, from universe-averaged models.
We note, however, that 
the distance dependence of the peculiar proper motion power spectrum will not be detected 
using the $\it Gaia$-Cosmicflows sample. When $\it Gaia$-like noise is added to the proper motions,
the power per distance bin is not significantly detected.  

The secular parallax dipole power is $\sqrt{P_{E1}} =226 ~\mu$as yr$^{-1}$ Mpc. 	
While secular parallax decreases as $1/D$, the normalized power (which is a proxy for the solar system's 
linear velocity) is not a function of distance. Secular parallax is therefore dominant for distances
$\gtrsim 40$ Mpc compared to the peculiar dipole. However, {\it Gaia} will not be sensitive 
enough to significantly detect secular parallax at that distance. 

\subsection{Impact on Distance-Independent Measurements}\label{sec:higher_modes}

\begin{figure}[t!]
    \centering
    \includegraphics[width=\columnwidth]{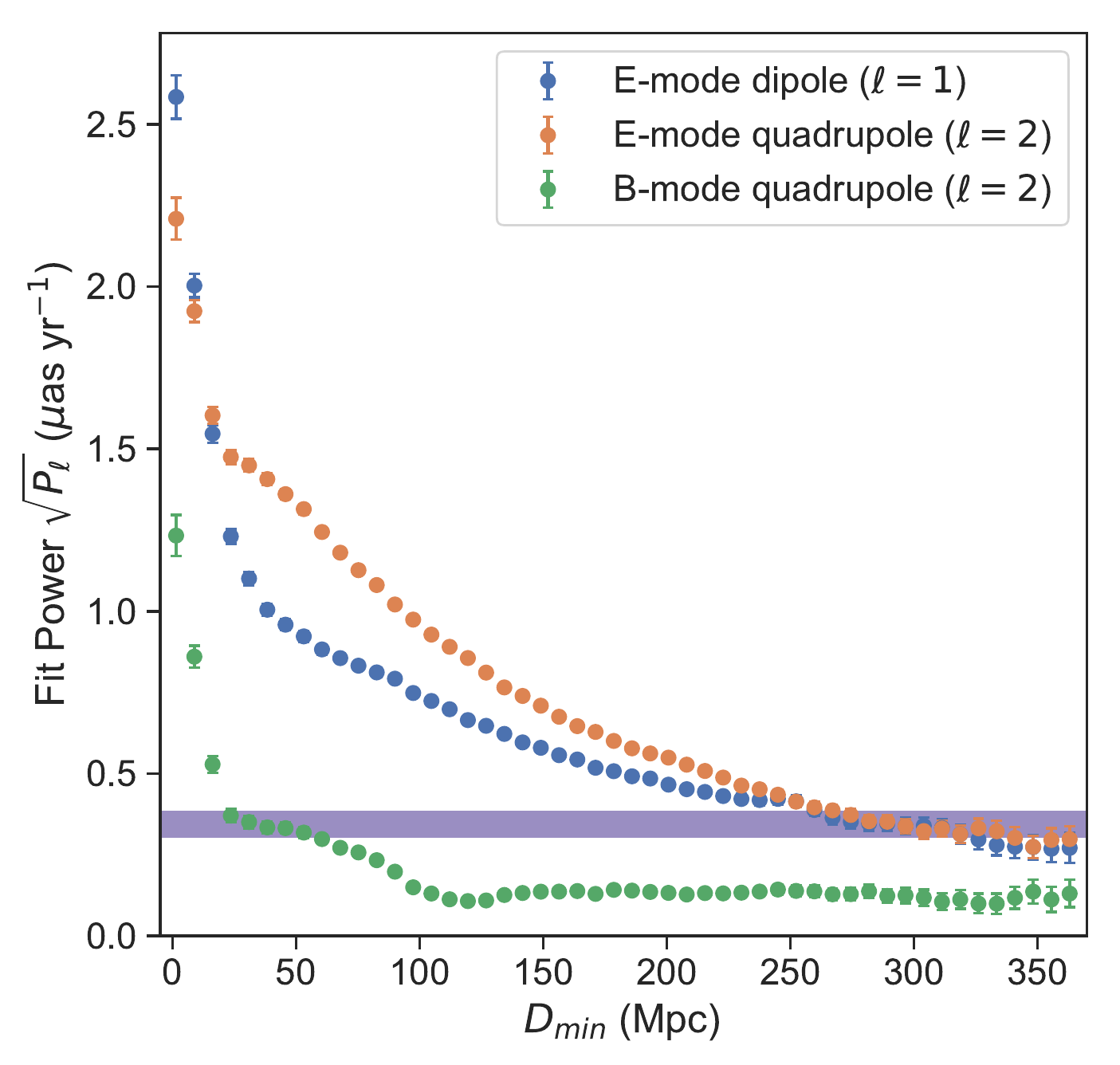}
    \caption{Distance-independent E-mode dipole (blue) and quadrupole (orange), and 
    B-mode quadrupole (green) fits to the peculiar 
    proper motions for varying minimum distances. The purple bar at 0.34 $\mu$as yr$^{-1}$
    represents the approximate total $\sqrt{P_2}$ (E2$+$B2) for $D_{min}>300$ Mpc cuts.  }
    \label{fig:vpec_quads}
\end{figure}

Cosmological effects such as the isotropy of the Hubble expansion and primordial 
gravitational waves induce distance-indepedent proper motion signals at low $\ell$. 
Here we assess the impact that peculiar velocities will have on distance-independent 
proper motion quadrupole measurements. In comparison to the sample of local galaxies 
selected for this work, distance independent measures may be made for quasars at
generally higher redshift. It is thus most relevant to compare the quadrupole powers recovered
for large distance cuts, where we expect the LSS power spectrum to be dominated by higher $\ell$. 
Figure~\ref{fig:vpec_quads} shows the results of simultaneously fitting 
an E-mode dipole and both E and B-mode quadrupoles to the noise-free peculiar proper
motions for varying $D_{min}$. Note that these are distance-independent fits, so the square root powers 
are 1-2 orders of magnitude lower than those in Figure~\ref{fig:vpec_E1E2E3} and cannot 
be simply scaled to a relative velocity. 
The powers in each mode approach minimum values $\sqrt{P_{\ell}}<0.4~\mu$as yr$^{-1}$ for large distance cuts. 
Below, we describe the implications of these fit powers for measurements of the isotropy 
of the Hubble expansion and gravitational waves. 

\subsubsection{Anisotropic Expansion}

Triaxial anisotropic expansion would be observable as an E-mode proper motion quadrupole, 
where galaxies appear to stream towards the directions of faster expansion \citep{Darling2014}.
The anisotropy is characterized by the fractional departure from the Hubble expansion in a given 
direction: 
$$ \Sigma_{i} = \frac{H_{i,0}}{H_0}-1,$$
where $i$ denotes the triaxial expansion axes $x$, $y$, and $z$. 
From Equations A1 and A2 of \citet{Darling2014}, we derive an expression to relate 
the total anisotropy to a proper motion quadrupole power by 
\begin{equation}
\sqrt{P_{E_2}} = H_0  \sqrt{\frac{4\pi}{5} \left(\Sigma_{x}^2 + \Sigma_{y}^2 + \Sigma_{z}^2\right)}, 
\end{equation}
where $H_0$ is the Hubble constant expressed in units of proper motion, which we 
assume to be $H_0 \approx 70$ km s$^{-1}$ Mpc$^{-1} = 15~\mu$as yr$^{-1}$. 

The mean peculiar E-mode $\ell=2$ power for $D_{min}>300$ Mpc is
$\sqrt{P_{E_2}} = 0.3~\mu$as yr$^{-1}$, or 0.014  in units of $H_0 \sqrt{4\pi/5}$. 
For each fit, we also solve for $\Sigma_{i}$ using the relations to the quadrupole coefficients 
derived in \cite{Darling2014}. The average maximum anisotropy inferred from the quadrupole
fits for $D_{min}>300$ is $\Sigma_{max} \approx 0.01$. Anisotropic expansion would therefore
be indistinguishable from LSS proper motions for anisotropy $<1\%$. 
The predicted end-of-mission anisotropy noise floor 
for {\it Gaia} quasars is about 2\% (roughly $\sqrt{P_{E2}} = 1 ~\mu$as yr$^{-1}$; \citealt{Paine2018}), 
so the anticipated limit on anisotropy from $\it Gaia$ is unaffected by LSS. 
However, improved astrometric precision from a next generation Very Large Array (ngVLA) or 
future space based astrometry mission may enable quadrupole measurements with quasars 
that will approach or be limited by the LSS power. 

\subsubsection{Gravitational Waves}
	
The primordial gravitational wave background will cause angular deflection of light rays 
with equal power in the E and B-modes for $\ell \geq 2$ \citep{Gwinn1997,Book2011}, which 
may be observable as extragalactic proper motions. The power in $\ell =2$ contributes $5/6$ of the total
gravitational wave signal, so the gravitational wave energy density may be estimated by 
\begin{equation}
\Omega_{GW} = \frac{6}{5} \frac{1}{4\pi} \frac{P_2}{H_0^2},
\end{equation}
where $P_2$ is the total quadrupole power (E and B-modes). 
\cite{Darling2018} found astrometric limits on the gravitational wave energy density 
of $\Omega_{GW}<0.0064$ for proper motions of 711 radio sources and
$\Omega_{GW}<0.011$ for proper motions of 508 radio sources combined with one 
epoch of {\it Gaia} data from the first data release. Additionally, they 
predicted 	a $\Omega_{GW} = 4\times 10^{-4}$ noise floor for {\it Gaia} 
end-of-mission AGN proper motions. 
	
For $D_{min}>300$ the total quadrupole square root power approaches $\sqrt{P_2}= 0.34~\mu$as yr$^{-1}$, 
which corresponds to $\Omega_{GW} = 5\times 10^{-4}$, similar to the predicted 
{\it Gaia} noise floor. However, the power in B2 is significantly lower than 
that in E2 for the peculiar proper motions, whereas gravitational waves produce equal power 
in E and B-modes. With improved astrometric precision, one may 
differentiate between the LSS quadrupole and gravitational waves by comparing the 
powers detected in the E and B-modes.

\section{Discussion and Conclusions}\label{sec:discussion}

The systematic errors of {\it Gaia} DR2 proper motions impose a noise floor that 
limits the amplitude of proper motion signals that may be studied with vector
spherical harmonic decomposition. 
The large-scale systematics have been  studied in detail by \cite{Lindegren2018}  and
\cite{Mignard2018}, and correspond to $\sim 40~\mu\rm{as~yr}^{-1}$ for large angular scales.
We find that the addition of distance data utilized in the secular parallax measurement, which
one would not expect to be spatially correlated, does not allow us to probe signals
below the systematic noise floor. We therefore find a 3500 $\mu$as yr$^{-1}$ Mpc upper
limit on the extragalactic secular parallax amplitude. Increased sample size would improve
the sensitivity of the measurement since uncertainties scale as $N^{-1/2}$. However, 
achieving the statistical uncertainty to detect the 80 $\mu$as yr$^{-1}$ Mpc
signal of secular parallax using {\it Gaia} DR2 would require a sample size of $\sim 10^6$, 
which is more than the number of galaxies in the local volume. 

Using the expected improvements to the sample size and proper motion uncertainties
of {\it Gaia}-Cosmicflows galaxies in future data releases, we predict a significant 
(5-10$\sigma$) detection of the secular parallax amplitude. This prediction is dependent
on the assumption that the systematic errors in {\it Gaia} DR2 will be resolved. 
The detection is also highly sensitive to sample selection: relatively bright, local 
galaxies must be included to significantly detect secular parallax. 
We find that the secular parallax is only detected for simulations with $D_{min}$ cuts 
$<5$ Mpc and $G$ magnitude cuts $<15$. The significance of the detection may also 
be improved by employing a proper motion amplitude cut $<150~\mu$as yr$^{-1}$, which 
limits the portion of the sample with highly noise-dominated proper motions. 
Very distant galaxies with expected secular parallaxes well below {\it Gaia}'s expected 
proper motion precision may be included in the sample, but have little to no effect on
the detected dipole as long as the signal is detected for local galaxies.  
{\it Gaia} data release 3 is expected to include results
for extended objects and quasars, which may have implications for the secular parallax
measurement. 

The ultimate goal of a secular parallax measurement would be extragalactic distance estimates.
However, the secular parallax may only be measured if a prior is assumed for the peculiar motion and distance,
such as those provided by Cosmicflows-3, and therefore an independent distance estimate from
secular parallax is most likely not possible for individual galaxies. 
Regardless, a statistical measurement of the secular parallax amplitude would lead to a constraint on the Hubble constant. 
If redshift is used as proxy for distance in Equation~\ref{eqn:parallax}, then the amplitude 
of the fit is a product of the secular parallax proper motion amplitude and the local value 
of the Hubble parameter, $H_0$. One therefore needs to either constrain the secular parallax
amplitude or assume the value from the CMB to infer the Hubble parameter. 
Our prediction does not directly translate to a redshift based fit because the galaxies 
closer than 5 Mpc that dominate the simulated fits most likely have large peculiar velocities
relative to the recessional velocities predicted by the Hubble flow. 
However, the fit for nearby galaxies will be useful to assess the mixing of the secular 
parallax and peculiar dipole signals.
From the results in Figure~\ref{fig:r_cuts}, we predict $\sim 2\sigma$ amplitude fits for
$D_{min}$ cuts $>20$ Mpc, and thus a $\sim 2\sigma$ upper limit on $H_0$ if the
secular parallax amplitude is assumed from the CMB dipole. 

The primary limitation of the predicted detection and any resulting constraint on the Hubble
constant is the dependence on very nearby galaxy 
proper motions, and thus the mixing of the secular parallax signal with the peculiar dipole. 
A potential method to improve the signal to noise 
at larger distances is to search for galaxy cluster members detected by {\it Gaia}. 
Individual proper motions could be averaged to determine the bulk motion of the cluster.
Though cluster peculiar velocities are $\sim500$ km s$^{-1}$\citep{Aghanim2018}, the peculiar dipole would
be lower in amplitude than the secular parallax for clusters at distances $\gtrsim 40$.
After {\it Gaia}, astrometry from ngVLA or a future space based mission 
with lower per object uncertainties would improve the signal measured from more distant 
galaxies. Though an ngVLA quasar survey would miss many local galaxies
observed by {\it Gaia}, the diminished peculiar dipole signal at the distances of relatively 
low redshift quasars may result in an improved constraint on the direction and amplitude of the 
secular parallax dipole. 

Previous works have also made predictions of the secular parallax measurement with
{\it Gaia}, namely,  \citet{Ding2009} and \citet{Hall2018}. Contrary to our results, 
neither work predicts a significant detection of the secular parallax. The discrepancy
with our results are most likely due to differences in the samples used for predictions. 
Both \citet{Ding2009} and \citet{Hall2018} assume larger samples of more distant objects,
whereas we find that significant secular parallax detection with {\it Gaia} is contingent on
galaxies closer than $\sim 5$ Mpc. Our simulations excluding galaxies closer than 
at least 20 Mpc are consistent with previous works. 

Our analysis of the Cosmicflows peculiar velocities are largely consistent 
with the LSS predictions of \citet{Hall2018} though our methods of simulating the peculiar 
velocities are independent. The noise-free, Cosmicflows-based peculiar proper motions 
show a dominant dipole for low distances bins and power generally shifting to higher $\ell$ 
modes with increasing distance. We find that the peculiar proper motion dipole
may be significantly detected with {\it Gaia} if one first detects or assumes the secular parallax
dipole from the CMB. However, the distance dependence of the peculiar 
proper motion power spectrum will not be detectable with {\it Gaia}.

We demonstrated that the peculiar velocities also produce low multipole correlated proper
motions that are distance independent, and therefore may impact other cosmological
proper motion measurements made with quasars for signals of $\sim 0.3 ~\mu$as yr$^{-1}$. 
This would be indistinguishable from $\sim 1\%$ anisotropic expansion and would dominate
the E-mode fields caused by a gravitational wave background. However, the gravitational 
wave energy density could still be constrained by studying the corresponding B-modes. 

\acknowledgments
We thank the anonymous referee for very helpful comments and corrections.
The authors acknowledge support from the NSF Graduate Research Fellowship Program
under grant DGE-1650115, the NSF grant AST-1411605, and the NASA grant 14-ATP14-0086.
RG acknowledges support from the European Research Council (ERC) under the European Union’s 
Horizon 2020 research and innovation programme (grant agreement no759194 - USNAC).
This work has made use of data from the European Space Agency (ESA)
mission {\it Gaia} (\url{https://www.cosmos.esa.int/gaia}), processed by
the {\it Gaia} Data Processing and Analysis Consortium (DPAC,
\url{https://www.cosmos.esa.int/web/gaia/dpac/consortium}). Funding
for the DPAC has been provided by national institutions, in particular
the institutions participating in the {\it Gaia} Multilateral Agreement.

\software{astropy \citep{astropy}, pyGaia,  TOPCAT \citep{topcat}
          }

\bibliography{references}

\begin{thebibliography}{}
\expandafter\ifx\csname natexlab\endcsname\relax\def\natexlab#1{#1}\fi
\providecommand{\url}[1]{\href{#1}{#1}}
\providecommand{\dodoi}[1]{doi:~\href{http://doi.org/#1}{\nolinkurl{#1}}}
\providecommand{\doeprint}[1]{\href{http://ascl.net/#1}{\nolinkurl{http://ascl.net/#1}}}
\providecommand{\doarXiv}[1]{\href{https://arxiv.org/abs/#1}{\nolinkurl{https://arxiv.org/abs/#1}}}

\bibitem[{{Astropy Collaboration} {et~al.}(2013){Astropy Collaboration},
  {Robitaille}, {Tollerud}, {Greenfield}, {Droettboom}, {Bray}, {Aldcroft},
  {Davis}, {Ginsburg}, {Price-Whelan}, {Kerzendorf}, {Conley}, {Crighton},
  {Barbary}, {Muna}, {Ferguson}, {Grollier}, {Parikh}, {Nair}, {Unther},
  {Deil}, {Woillez}, {Conseil}, {Kramer}, {Turner}, {Singer}, {Fox}, {Weaver},
  {Zabalza}, {Edwards}, {Azalee Bostroem}, {Burke}, {Casey}, {Crawford},
  {Dencheva}, {Ely}, {Jenness}, {Labrie}, {Lim}, {Pierfederici}, {Pontzen},
  {Ptak}, {Refsdal}, {Servillat}, \& {Streicher}}]{astropy}
{Astropy Collaboration}, {Robitaille}, T.~P., {Tollerud}, E.~J., {et~al.} 2013,
  \aap, 558, A33, \dodoi{10.1051/0004-6361/201322068}

\bibitem[{{Bachchan} {et~al.}(2016){Bachchan}, {Hobbs}, \&
  {Lindegren}}]{Bachchan2016}
{Bachchan}, R.~K., {Hobbs}, D., \& {Lindegren}, L. 2016, \aap, 589, A71,
  \dodoi{10.1051/0004-6361/201527935}

\bibitem[{{Book} \& {Flanagan}(2011)}]{Book2011}
{Book}, L.~G., \& {Flanagan}, {\'E}.~{\'E}. 2011, \prd, 83, 024024,
  \dodoi{10.1103/PhysRevD.83.024024}

\bibitem[{{Darling}(2014)}]{Darling2014}
{Darling}, J. 2014, \mnras, 442, L66, \dodoi{10.1093/mnrasl/slu057}

\bibitem[{{Darling} \& {Truebenbach}(2018)}]{Darling&Truebenbach2018}
{Darling}, J., \& {Truebenbach}, A.~E. 2018, \apj, 864, 37,
  \dodoi{10.3847/1538-4357/aad3d0}

\bibitem[{{Darling} {et~al.}(2018){Darling}, {Truebenbach}, \&
  {Paine}}]{Darling2018}
{Darling}, J., {Truebenbach}, A.~E., \& {Paine}, J. 2018, \apj, 861, 113,
  \dodoi{10.3847/1538-4357/aac772}

\bibitem[{{Ding} \& {Croft}(2009)}]{Ding2009}
{Ding}, F., \& {Croft}, R. A.~C. 2009, \mnras, 397, 1739,
  \dodoi{10.1111/j.1365-2966.2009.15111.x}

\bibitem[{{Djorgovski} \& {Davis}(1987)}]{Djorgovski1987}
{Djorgovski}, S., \& {Davis}, M. 1987, \apj, 313, 59, \dodoi{10.1086/164948}

\bibitem[{{Dressler} {et~al.}(1987){Dressler}, {Lynden-Bell}, {Burstein},
  {Davies}, {Faber}, {Terlevich}, \& {Wegner}}]{Dressler1987}
{Dressler}, A., {Lynden-Bell}, D., {Burstein}, D., {et~al.} 1987, \apj, 313,
  42, \dodoi{10.1086/164947}

\bibitem[{{Gaia Collaboration} {et~al.}(2016){Gaia Collaboration}, {Prusti},
  {de Bruijne}, {Brown}, {Vallenari}, {Babusiaux}, {Bailer-Jones}, {Bastian},
  {Biermann}, {Evans}, {Eyer}, {Jansen}, {Jordi}, {Klioner}, {Lammers},
  {Lindegren}, {Luri}, {Mignard}, {Milligan}, {Panem}, {Poinsignon},
  {Pourbaix}, {Randich}, {Sarri}, {Sartoretti}, {Siddiqui}, {Soubiran},
  {Valette}, {van Leeuwen}, {Walton}, {Aerts}, {Arenou}, {Cropper}, {Drimmel},
  {H{\o}g}, {Katz}, {Lattanzi}, {O'Mullane}, {Grebel}, {Holland}, {Huc},
  {Passot}, {Bramante}, {Cacciari}, {Casta{\~n}eda}, {Chaoul}, {Cheek}, {De
  Angeli}, {Fabricius}, {Guerra}, {Hern{\'a}ndez}, {Jean-Antoine-Piccolo},
  {Masana}, {Messineo}, {Mowlavi}, {Nienartowicz}, {Ord{\'o}{\~n}ez-Blanco},
  {Panuzzo}, {Portell}, {Richards}, {Riello}, {Seabroke}, {Tanga},
  {Th{\'e}venin}, {Torra}, {Els}, {Gracia-Abril}, {Comoretto},
  {Garcia-Reinaldos}, {Lock}, {Mercier}, {Altmann}, {Andrae}, {Astraatmadja},
  {Bellas-Velidis}, {Benson}, {Berthier}, {Blomme}, {Busso}, {Carry},
  {Cellino}, {Clementini}, {Cowell}, {Creevey}, {Cuypers}, {Davidson}, {De
  Ridder}, {de Torres}, {Delchambre}, {Dell'Oro}, {Ducourant}, {Fr{\'e}mat},
  {Garc{\'\i}a-Torres}, {Gosset}, {Halbwachs}, {Hambly}, {Harrison}, {Hauser},
  {Hestroffer}, {Hodgkin}, {Huckle}, {Hutton}, {Jasniewicz}, {Jordan},
  {Kontizas}, {Korn}, {Lanzafame}, {Manteiga}, {Moitinho}, {Muinonen},
  {Osinde}, {Pancino}, {Pauwels}, {Petit}, {Recio-Blanco}, {Robin}, {Sarro},
  {Siopis}, {Smith}, {Smith}, {Sozzetti}, {Thuillot}, {van Reeven}, {Viala},
  {Abbas}, {Abreu Aramburu}, {Accart}, {Aguado}, {Allan}, {Allasia},
  {Altavilla}, {{\'A}lvarez}, {Alves}, {Anderson}, {Andrei}, {Anglada Varela},
  {Antiche}, {Antoja}, {Ant{\'o}n}, {Arcay}, {Atzei}, {Ayache}, {Bach},
  {Baker}, {Balaguer-N{\'u}{\~n}ez}, {Barache}, {Barata}, {Barbier}, {Barblan},
  {Baroni}, {Barrado y Navascu{\'e}s}, {Barros}, {Barstow}, {Becciani},
  {Bellazzini}, {Bellei}, {Bello Garc{\'\i}a}, {Belokurov}, {Bendjoya},
  {Berihuete}, {Bianchi}, {Bienaym{\'e}}, {Billebaud}, {Blagorodnova},
  {Blanco-Cuaresma}, {Boch}, {Bombrun}, {Borrachero}, {Bouquillon}, {Bourda},
  {Bouy}, {Bragaglia}, {Breddels}, {Brouillet}, {Br{\"u}semeister},
  {Bucciarelli}, {Budnik}, {Burgess}, {Burgon}, {Burlacu}, {Busonero}, {Buzzi},
  {Caffau}, {Cambras}, {Campbell}, {Cancelliere}, {Cantat-Gaudin}, {Carlucci},
  {Carrasco}, {Castellani}, {Charlot}, {Charnas}, {Charvet}, {Chassat},
  {Chiavassa}, {Clotet}, {Cocozza}, {Collins}, {Collins}, {Costigan}, {Crifo},
  {Cross}, {Crosta}, {Crowley}, {Dafonte}, {Damerdji}, {Dapergolas}, {David},
  {David}, {De Cat}, {de Felice}, {de Laverny}, {De Luise}, {De March}, {de
  Martino}, {de Souza}, {Debosscher}, {del Pozo}, {Delbo}, {Delgado},
  {Delgado}, {di Marco}, {Di Matteo}, {Diakite}, {Distefano}, {Dolding}, {Dos
  Anjos}, {Drazinos}, {Dur{\'a}n}, {Dzigan}, {Ecale}, {Edvardsson}, {Enke},
  {Erdmann}, {Escolar}, {Espina}, {Evans}, {Eynard Bontemps}, {Fabre},
  {Fabrizio}, {Faigler}, {Falc{\~a}o}, {Farr{\`a}s Casas}, {Faye}, {Federici},
  {Fedorets}, {Fern{\'a}ndez-Hern{\'a}ndez}, {Fernique}, {Fienga}, {Figueras},
  {Filippi}, {Findeisen}, {Fonti}, {Fouesneau}, {Fraile}, {Fraser}, {Fuchs},
  {Furnell}, {Gai}, {Galleti}, {Galluccio}, {Garabato}, {Garc{\'\i}a-Sedano},
  {Gar{\'e}}, {Garofalo}, {Garralda}, {Gavras}, {Gerssen}, {Geyer}, {Gilmore},
  {Girona}, {Giuffrida}, {Gomes}, {Gonz{\'a}lez-Marcos},
  {Gonz{\'a}lez-N{\'u}{\~n}ez}, {Gonz{\'a}lez-Vidal}, {Granvik}, {Guerrier},
  {Guillout}, {Guiraud}, {G{\'u}rpide}, {Guti{\'e}rrez-S{\'a}nchez}, {Guy},
  {Haigron}, {Hatzidimitriou}, {Haywood}, {Heiter}, {Helmi}, {Hobbs},
  {Hofmann}, {Holl}, {Holland }, {Hunt}, {Hypki}, {Icardi}, {Irwin}, {Jevardat
  de Fombelle}, {Jofr{\'e}}, {Jonker}, {Jorissen}, {Julbe}, {Karampelas},
  {Kochoska}, {Kohley}, {Kolenberg}, {Kontizas}, {Koposov}, {Kordopatis},
  {Koubsky}, {Kowalczyk}, {Krone-Martins}, {Kudryashova}, {Kull}, {Bachchan},
  {Lacoste-Seris}, {Lanza}, {Lavigne}, {Le Poncin-Lafitte}, {Lebreton},
  {Lebzelter}, {Leccia}, {Leclerc}, {Lecoeur-Taibi}, {Lemaitre}, {Lenhardt},
  {Leroux}, {Liao}, {Licata}, {Lindstr{\o}m}, {Lister}, {Livanou}, {Lobel},
  {L{\"o}ffler}, {L{\'o}pez}, {Lopez-Lozano}, {Lorenz}, {Loureiro},
  {MacDonald}, {Magalh{\~a}es Fernandes}, {Managau}, {Mann}, {Mantelet},
  {Marchal}, {Marchant}, {Marconi}, {Marie}, {Marinoni}, {Marrese},
  {Marschalk{\'o}}, {Marshall}, {Mart{\'\i}n-Fleitas}, {Martino}, {Mary},
  {Matijevi{\v{c}}}, {Mazeh}, {McMillan}, {Messina}, {Mestre}, {Michalik},
  {Millar}, {Miranda}, {Molina}, {Molinaro}, {Molinaro}, {Moln{\'a}r},
  {Moniez}, {Montegriffo}, {Monteiro}, {Mor}, {Mora}, {Morbidelli}, {Morel},
  {Morgenthaler}, {Morley}, {Morris}, {Mulone}, {Muraveva}, {Musella},
  {Narbonne}, {Nelemans}, {Nicastro}, {Noval}, {Ord{\'e}novic},
  {Ordieres-Mer{\'e}}, {Osborne}, {Pagani}, {Pagano}, {Pailler}, {Palacin},
  {Palaversa}, {Parsons}, {Paulsen}, {Pecoraro}, {Pedrosa}, {Pentik{\"a}inen},
  {Pereira}, {Pichon}, {Piersimoni}, {Pineau}, {Plachy}, {Plum}, {Poujoulet},
  {Pr{\v{s}}a}, {Pulone}, {Ragaini}, {Rago}, {Rambaux}, {Ramos-Lerate},
  {Ranalli}, {Rauw}, {Read}, {Regibo}, {Renk}, {Reyl{\'e}}, {Ribeiro},
  {Rimoldini}, {Ripepi}, {Riva}, {Rixon}, {Roelens}, {Romero-G{\'o}mez},
  {Rowell}, {Royer}, {Rudolph}, {Ruiz-Dern}, {Sadowski}, {Sagrist{\`a}
  Sell{\'e}s}, {Sahlmann}, {Salgado}, {Salguero}, {Sarasso}, {Savietto},
  {Schnorhk}, {Schultheis}, {Sciacca}, {Segol}, {Segovia}, {Segransan},
  {Serpell}, {Shih}, {Smareglia}, {Smart}, {Smith}, {Solano}, {Solitro},
  {Sordo}, {Soria Nieto}, {Souchay}, {Spagna}, {Spoto}, {Stampa}, {Steele},
  {Steidelm{\"u}ller}, {Stephenson}, {Stoev}, {Suess}, {S{\"u}veges}, {Surdej},
  {Szabados}, {Szegedi-Elek}, {Tapiador}, {Taris}, {Tauran}, {Taylor},
  {Teixeira}, {Terrett}, {Tingley}, {Trager}, {Turon}, {Ulla}, {Utrilla},
  {Valentini}, {van Elteren}, {Van Hemelryck}, {van Leeuwen}, {Varadi},
  {Vecchiato}, {Veljanoski}, {Via}, {Vicente}, {Vogt}, {Voss}, {Votruba},
  {Voutsinas}, {Walmsley}, {Weiler}, {Weingrill}, {Werner}, {Wevers},
  {Whitehead}, {Wyrzykowski}, {Yoldas}, {{\v{Z}}erjal}, {Zucker}, {Zurbach},
  {Zwitter}, {Alecu}, {Allen}, {Allende Prieto}, {Amorim},
  {Anglada-Escud{\'e}}, {Arsenijevic}, {Azaz}, {Balm}, {Beck}, {Bernstein},
  {Bigot}, {Bijaoui}, {Blasco}, {Bonfigli}, {Bono}, {Boudreault}, {Bressan},
  {Brown}, {Brunet}, {Bunclark}, {Buonanno}, {Butkevich}, {Carret}, {Carrion},
  {Chemin}, {Ch{\'e}reau}, {Corcione}, {Darmigny}, {de Boer}, {de Teodoro}, {de
  Zeeuw}, {Delle Luche}, {Domingues}, {Dubath}, {Fodor}, {Fr{\'e}zouls},
  {Fries}, {Fustes}, {Fyfe}, {Gallardo}, {Gallegos}, {Gardiol}, {Gebran},
  {Gomboc}, {G{\'o}mez}, {Grux}, {Gueguen}, {Heyrovsky}, {Hoar}, {Iannicola},
  {Isasi Parache}, {Janotto}, {Joliet}, {Jonckheere}, {Keil}, {Kim},
  {Klagyivik}, {Klar}, {Knude}, {Kochukhov}, {Kolka}, {Kos}, {Kutka}, {Lainey},
  {LeBouquin}, {Liu}, {Loreggia}, {Makarov}, {Marseille}, {Martayan},
  {Martinez-Rubi}, {Massart}, {Meynadier}, {Mignot}, {Munari}, {Nguyen},
  {Nordlander}, {Ocvirk}, {O'Flaherty}, {Olias Sanz}, {Ortiz}, {Osorio},
  {Oszkiewicz}, {Ouzounis}, {Palmer}, {Park}, {Pasquato}, {Peltzer}, {Peralta},
  {P{\'e}turaud}, {Pieniluoma}, {Pigozzi}, {Poels}, {Prat}, {Prod'homme},
  {Raison}, {Rebordao}, {Risquez}, {Rocca-Volmerange}, {Rosen}, {Ruiz-Fuertes},
  {Russo}, {Sembay}, {Serraller Vizcaino}, {Short}, {Siebert}, {Silva},
  {Sinachopoulos}, {Slezak}, {Soffel}, {Sosnowska}, {Strai{\v{z}}ys}, {ter
  Linden}, {Terrell}, {Theil}, {Tiede}, {Troisi}, {Tsalmantza}, {Tur},
  {Vaccari}, {Vachier}, {Valles}, {Van Hamme}, {Veltz}, {Virtanen}, {Wallut},
  {Wichmann}, {Wilkinson}, {Ziaeepour}, \& {Zschocke}}]{GaiaCollaboration2016}
{Gaia Collaboration}, {Prusti}, T., {de Bruijne}, J.~H.~J., {et~al.} 2016,
  \aap, 595, A1, \dodoi{10.1051/0004-6361/201629272}

\bibitem[{{Gaia Collaboration} {et~al.}(2018{\natexlab{a}}){Gaia
  Collaboration}, {Brown}, {Vallenari}, {Prusti}, {de Bruijne}, {Babusiaux},
  {Bailer-Jones}, {Biermann}, {Evans}, {Eyer}, {Jansen}, {Jordi}, {Klioner},
  {Lammers}, {Lindegren}, {Luri}, {Mignard}, {Panem}, {Pourbaix}, {Randich},
  {Sartoretti}, {Siddiqui}, {Soubiran}, {van Leeuwen}, {Walton}, {Arenou},
  {Bastian}, {Cropper}, {Drimmel}, {Katz}, {Lattanzi}, {Bakker}, {Cacciari},
  {Casta{\~n}eda}, {Chaoul}, {Cheek}, {De Angeli}, {Fabricius}, {Guerra},
  {Holl}, {Masana}, {Messineo}, {Mowlavi}, {Nienartowicz}, {Panuzzo},
  {Portell}, {Riello}, {Seabroke}, {Tanga}, {Th{\'e}venin}, {Gracia-Abril},
  {Comoretto}, {Garcia-Reinaldos}, {Teyssier}, {Altmann}, {Andrae}, {Audard},
  {Bellas-Velidis}, {Benson}, {Berthier}, {Blomme}, {Burgess}, {Busso},
  {Carry}, {Cellino}, {Clementini}, {Clotet}, {Creevey}, {Davidson}, {De
  Ridder}, {Delchambre}, {Dell'Oro}, {Ducourant},
  {Fern{\'a}ndez-Hern{\'a}ndez}, {Fouesneau}, {Fr{\'e}mat}, {Galluccio},
  {Garc{\'\i}a-Torres}, {Gonz{\'a}lez-N{\'u}{\~n}ez}, {Gonz{\'a}lez-Vidal},
  {Gosset}, {Guy}, {Halbwachs}, {Hambly}, {Harrison}, {Hern{\'a}ndez},
  {Hestroffer}, {Hodgkin}, {Hutton}, {Jasniewicz}, {Jean-Antoine-Piccolo},
  {Jordan}, {Korn}, {Krone-Martins}, {Lanzafame}, {Lebzelter}, {L{\"o}ffler},
  {Manteiga}, {Marrese}, {Mart{\'\i}n-Fleitas}, {Moitinho}, {Mora}, {Muinonen},
  {Osinde}, {Pancino}, {Pauwels}, {Petit}, {Recio-Blanco}, {Richards},
  {Rimoldini}, {Robin}, {Sarro}, {Siopis}, {Smith}, {Sozzetti}, {S{\"u}veges},
  {Torra}, {van Reeven}, {Abbas}, {Abreu Aramburu}, {Accart}, {Aerts},
  {Altavilla}, {{\'A}lvarez}, {Alvarez}, {Alves}, {Anderson}, {Andrei},
  {Anglada Varela}, {Antiche}, {Antoja}, {Arcay}, {Astraatmadja}, {Bach},
  {Baker}, {Balaguer-N{\'u}{\~n}ez}, {Balm}, {Barache}, {Barata}, {Barbato},
  {Barblan}, {Barklem}, {Barrado}, {Barros}, {Barstow}, {Bartholom{\'e}
  Mu{\~n}oz}, {Bassilana}, {Becciani}, {Bellazzini}, {Berihuete}, {Bertone},
  {Bianchi}, {Bienaym{\'e}}, {Blanco-Cuaresma}, {Boch}, {Boeche}, {Bombrun},
  {Borrachero}, {Bossini}, {Bouquillon}, {Bourda}, {Bragaglia}, {Bramante},
  {Breddels}, {Bressan}, {Brouillet}, {Br{\"u}semeister}, {Brugaletta},
  {Bucciarelli}, {Burlacu}, {Busonero}, {Butkevich}, {Buzzi}, {Caffau},
  {Cancelliere}, {Cannizzaro}, {Cantat-Gaudin}, {Carballo}, {Carlucci},
  {Carrasco}, {Casamiquela}, {Castellani}, {Castro-Ginard}, {Charlot},
  {Chemin}, {Chiavassa}, {Cocozza}, {Costigan}, {Cowell}, {Crifo}, {Crosta},
  {Crowley}, {Cuypers}, {Dafonte}, {Damerdji}, {Dapergolas}, {David}, {David},
  {de Laverny}, {De Luise}, {De March}, {de Martino}, {de Souza}, {de Torres},
  {Debosscher}, {del Pozo}, {Delbo}, {Delgado}, {Delgado}, {Di Matteo},
  {Diakite}, {Diener}, {Distefano}, {Dolding}, {Drazinos}, {Dur{\'a}n},
  {Edvardsson}, {Enke}, {Eriksson}, {Esquej}, {Eynard Bontemps}, {Fabre},
  {Fabrizio}, {Faigler}, {Falc{\~a}o}, {Farr{\`a}s Casas}, {Federici},
  {Fedorets}, {Fernique}, {Figueras}, {Filippi}, {Findeisen}, {Fonti},
  {Fraile}, {Fraser}, {Fr{\'e}zouls}, {Gai}, {Galleti}, {Garabato},
  {Garc{\'\i}a-Sedano}, {Garofalo}, {Garralda}, {Gavel}, {Gavras}, {Gerssen},
  {Geyer}, {Giacobbe}, {Gilmore}, {Girona}, {Giuffrida}, {Glass}, {Gomes},
  {Granvik}, {Gueguen}, {Guerrier}, {Guiraud}, {Guti{\'e}rrez-S{\'a}nchez},
  {Haigron}, {Hatzidimitriou}, {Hauser}, {Haywood}, {Heiter}, {Helmi}, {Heu},
  {Hilger}, {Hobbs}, {Hofmann}, {Holland}, {Huckle}, {Hypki}, {Icardi},
  {Jan{\ss}en}, {Jevardat de Fombelle}, {Jonker}, {Juh{\'a}sz}, {Julbe},
  {Karampelas}, {Kewley}, {Klar}, {Kochoska}, {Kohley}, {Kolenberg},
  {Kontizas}, {Kontizas}, {Koposov}, {Kordopatis}, {Kostrzewa-Rutkowska},
  {Koubsky}, {Lambert}, {Lanza}, {Lasne}, {Lavigne}, {Le Fustec}, {Le
  Poncin-Lafitte}, {Lebreton}, {Leccia}, {Leclerc}, {Lecoeur-Taibi},
  {Lenhardt}, {Leroux}, {Liao}, {Licata}, {Lindstr{\o}m}, {Lister}, {Livanou},
  {Lobel}, {L{\'o}pez}, {Managau}, {Mann}, {Mantelet}, {Marchal}, {Marchant},
  {Marconi}, {Marinoni}, {Marschalk{\'o}}, {Marshall}, {Martino}, {Marton},
  {Mary}, {Massari}, {Matijevi{\v{c}}}, {Mazeh}, {McMillan}, {Messina},
  {Michalik}, {Millar}, {Molina}, {Molinaro}, {Moln{\'a}r}, {Montegriffo},
  {Mor}, {Morbidelli}, {Morel}, {Morris}, {Mulone}, {Muraveva}, {Musella},
  {Nelemans}, {Nicastro}, {Noval}, {O'Mullane}, {Ord{\'e}novic},
  {Ord{\'o}{\~n}ez-Blanco}, {Osborne}, {Pagani}, {Pagano}, {Pailler},
  {Palacin}, {Palaversa}, {Panahi}, {Pawlak}, {Piersimoni}, {Pineau}, {Plachy},
  {Plum}, {Poggio}, {Poujoulet}, {Pr{\v{s}}a}, {Pulone}, {Racero}, {Ragaini},
  {Rambaux}, {Ramos-Lerate}, {Regibo}, {Reyl{\'e}}, {Riclet}, {Ripepi}, {Riva},
  {Rivard}, {Rixon}, {Roegiers}, {Roelens}, {Romero-G{\'o}mez}, {Rowell},
  {Royer}, {Ruiz-Dern}, {Sadowski}, {Sagrist{\`a} Sell{\'e}s}, {Sahlmann},
  {Salgado}, {Salguero}, {Sanna}, {Santana-Ros}, {Sarasso}, {Savietto},
  {Schultheis}, {Sciacca}, {Segol}, {Segovia}, {S{\'e}gransan}, {Shih},
  {Siltala}, {Silva}, {Smart}, {Smith}, {Solano}, {Solitro}, {Sordo}, {Soria
  Nieto}, {Souchay}, {Spagna}, {Spoto}, {Stampa}, {Steele},
  {Steidelm{\"u}ller}, {Stephenson}, {Stoev}, {Suess}, {Surdej}, {Szabados},
  {Szegedi-Elek}, {Tapiador}, {Taris}, {Tauran}, {Taylor}, {Teixeira},
  {Terrett}, {Teyssand ier}, {Thuillot}, {Titarenko}, {Torra Clotet}, {Turon},
  {Ulla}, {Utrilla}, {Uzzi}, {Vaillant}, {Valentini}, {Valette}, {van Elteren},
  {Van Hemelryck}, {van Leeuwen}, {Vaschetto}, {Vecchiato}, {Veljanoski},
  {Viala}, {Vicente}, {Vogt}, {von Essen}, {Voss}, {Votruba}, {Voutsinas},
  {Walmsley}, {Weiler}, {Wertz}, {Wevers}, {Wyrzykowski}, {Yoldas},
  {{\v{Z}}erjal}, {Ziaeepour}, {Zorec}, {Zschocke}, {Zucker}, {Zurbach}, \&
  {Zwitter}}]{Brown2018}
{Gaia Collaboration}, {Brown}, A.~G.~A., {Vallenari}, A., {et~al.}
  2018{\natexlab{a}}, \aap, 616, A1, \dodoi{10.1051/0004-6361/201833051}

\bibitem[{{Gaia Collaboration} {et~al.}(2018{\natexlab{b}}){Gaia
  Collaboration}, {Mignard}, {Klioner}, {Lindegren}, {Hern{\'a}ndez},
  {Bastian}, {Bombrun}, {Hobbs}, {Lammers}, {Michalik}, {Ramos-Lerate},
  {Biermann}, {Fern{\'a}ndez-Hern{\'a}ndez}, {Geyer}, {Hilger}, {Siddiqui},
  {Steidelm{\"u}ller}, {Babusiaux}, {Barache}, {Lambert}, {Andrei}, {Bourda},
  {Charlot}, {Brown}, {Vallenari}, {Prusti}, {de Bruijne}, {Bailer-Jones},
  {Evans}, {Eyer}, {Jansen}, {Jordi}, {Luri}, {Panem}, {Pourbaix}, {Randich},
  {Sartoretti}, {Soubiran}, {van Leeuwen}, {Walton}, {Arenou}, {Cropper},
  {Drimmel}, {Katz}, {Lattanzi}, {Bakker}, {Cacciari}, {Casta{\~n}eda},
  {Chaoul}, {Cheek}, {De Angeli}, {Fabricius}, {Guerra}, {Holl}, {Masana},
  {Messineo}, {Mowlavi}, {Nienartowicz}, {Panuzzo}, {Portell}, {Riello},
  {Seabroke}, {Tanga}, {Th{\'e}venin}, {Gracia-Abril}, {Comoretto},
  {Garcia-Reinaldos}, {Teyssier}, {Altmann}, {Andrae}, {Audard},
  {Bellas-Velidis}, {Benson}, {Berthier}, {Blomme}, {Burgess}, {Busso},
  {Carry}, {Cellino}, {Clementini}, {Clotet}, {Creevey}, {Davidson}, {De
  Ridder}, {Delchambre}, {Dell'Oro}, {Ducourant}, {Fouesneau}, {Fr{\'e}mat},
  {Galluccio}, {Garc{\'\i}a-Torres}, {Gonz{\'a}lez-N{\'u}{\~n}ez},
  {Gonz{\'a}lez-Vidal}, {Gosset}, {Guy}, {Halbwachs}, {Hambly}, {Harrison},
  {Hestroffer}, {Hodgkin}, {Hutton}, {Jasniewicz}, {Jean-Antoine-Piccolo},
  {Jordan}, {Korn}, {Krone-Martins}, {Lanzafame}, {Lebzelter}, {L{\"o}ffler},
  {Manteiga}, {Marrese}, {Mart{\'\i}n-Fleitas}, {Moitinho}, {Mora}, {Muinonen},
  {Osinde}, {Pancino}, {Pauwels}, {Petit}, {Recio-Blanco}, {Richards},
  {Rimoldini}, {Robin}, {Sarro}, {Siopis}, {Smith}, {Sozzetti}, {S{\"u}veges},
  {Torra}, {van Reeven}, {Abbas}, {Abreu Aramburu}, {Accart}, {Aerts},
  {Altavilla}, {{\'A}lvarez}, {Alvarez}, {Alves}, {Anderson}, {Anglada Varela},
  {Antiche}, {Antoja}, {Arcay}, {Astraatmadja}, {Bach}, {Baker},
  {Balaguer-N{\'u}{\~n}ez}, {Balm}, {Barata}, {Barbato}, {Barblan}, {Barklem},
  {Barrado}, {Barros}, {Barstow}, {Bartholom{\'e} Mu{\~n}oz}, {Bassilana},
  {Becciani}, {Bellazzini}, {Berihuete}, {Bertone}, {Bianchi}, {Bienaym{\'e}},
  {Blanco-Cuaresma}, {Boch}, {Boeche}, {Borrachero}, {Bossini}, {Bouquillon},
  {Bragaglia}, {Bramante}, {Breddels}, {Bressan}, {Brouillet},
  {Br{\"u}semeister}, {Brugaletta}, {Bucciarelli}, {Burlacu}, {Busonero},
  {Butkevich}, {Buzzi}, {Caffau}, {Cancelliere}, {Cannizzaro}, {Cantat-Gaudin},
  {Carballo}, {Carlucci}, {Carrasco}, {Casamiquela}, {Castellani},
  {Castro-Ginard}, {Chemin}, {Chiavassa}, {Cocozza}, {Costigan}, {Cowell},
  {Crifo}, {Crosta}, {Crowley}, {Cuypers}, {Dafonte}, {Damerdji}, {Dapergolas},
  {David}, {David}, {de Laverny}, {De Luise}, {De March}, {de Souza}, {de
  Torres}, {Debosscher}, {del Pozo}, {Delbo}, {Delgado}, {Delgado}, {Diakite},
  {Diener}, {Distefano}, {Dolding}, {Drazinos}, {Dur{\'a}n}, {Edvardsson},
  {Enke}, {Eriksson}, {Esquej}, {Eynard Bontemps}, {Fabre}, {Fabrizio},
  {Faigler}, {Falc{\~a}o}, {Farr{\`a}s Casas}, {Federici}, {Fedorets},
  {Fernique}, {Figueras}, {Filippi}, {Findeisen}, {Fonti}, {Fraile}, {Fraser},
  {Fr{\'e}zouls}, {Gai}, {Galleti}, {Garabato}, {Garc{\'\i}a-Sedano},
  {Garofalo}, {Garralda}, {Gavel}, {Gavras}, {Gerssen}, {Giacobbe}, {Gilmore},
  {Girona}, {Giuffrida}, {Glass}, {Gomes}, {Granvik}, {Gueguen}, {Guerrier},
  {Guiraud}, {Guti{\'e}}, {Haigron}, {Hatzidimitriou}, {Hauser}, {Haywood},
  {Heiter}, {Helmi}, {Heu}, {Hofmann}, {Holland }, {Huckle}, {Hypki}, {Icardi},
  {Jan{\ss}en}, {Jevardat de Fombelle}, {Jonker}, {Juh{\'a}sz}, {Julbe},
  {Karampelas}, {Kewley}, {Klar}, {Kochoska}, {Kohley}, {Kolenberg},
  {Kontizas}, {Kontizas}, {Koposov}, {Kordopatis}, {Kostrzewa-Rutkowska},
  {Koubsky}, {Lanza}, {Lasne}, {Lavigne}, {Le Fustec}, {Le Poncin-Lafitte},
  {Lebreton}, {Leccia}, {Leclerc}, {Lecoeur-Taibi}, {Lenhardt}, {Leroux},
  {Liao}, {Licata}, {Lindstr{\o}m}, {Lister}, {Livanou}, {Lobel}, {L{\'o}pez},
  {Managau}, {Mann}, {Mantelet}, {Marchal}, {Marchant}, {Marconi}, {Marinoni},
  {Marschalk{\'o}}, {Marshall}, {Martino}, {Marton}, {Mary}, {Massari},
  {Matijevi{\v{c}}}, {Mazeh}, {McMillan}, {Messina}, {Millar}, {Molina},
  {Molinaro}, {Moln{\'a}r}, {Montegriffo}, {Mor}, {Morbidelli}, {Morel},
  {Morris}, {Mulone}, {Muraveva}, {Musella}, {Nelemans}, {Nicastro}, {Noval},
  {O'Mullane}, {Ord{\'e}novic}, {Ord{\'o}{\~n}ez-Blanco}, {Osborne}, {Pagani},
  {Pagano}, {Pailler}, {Palacin}, {Palaversa}, {Panahi}, {Pawlak},
  {Piersimoni}, {Pineau}, {Plachy}, {Plum}, {Poggio}, {Poujoulet},
  {Pr{\v{s}}a}, {Pulone}, {Racero}, {Ragaini}, {Rambaux}, {Regibo},
  {Reyl{\'e}}, {Riclet}, {Ripepi}, {Riva}, {Rivard}, {Rixon}, {Roegiers},
  {Roelens}, {Romero-G{\'o}mez}, {Rowell}, {Royer}, {Ruiz-Dern}, {Sadowski},
  {Sagrist{\`a} Sell{\'e}s}, {Sahlmann}, {Salgado}, {Salguero}, {Sanna},
  {Santana-Ros}, {Sarasso}, {Savietto}, {Schultheis}, {Sciacca}, {Segol},
  {Segovia}, {S{\'e}gransan}, {Shih}, {Siltala}, {Silva}, {Smart}, {Smith},
  {Solano}, {Solitro}, {Sordo}, {Soria Nieto}, {Souchay}, {Spagna}, {Spoto},
  {Stampa}, {Steele}, {Stephenson}, {Stoev}, {Suess}, {Surdej}, {Szabados},
  {Szegedi-Elek}, {Tapiador}, {Taris}, {Tauran}, {Taylor}, {Teixeira},
  {Terrett}, {Teyssand ier}, {Thuillot}, {Titarenko}, {Torra Clotet}, {Turon},
  {Ulla}, {Utrilla}, {Uzzi}, {Vaillant}, {Valentini}, {Valette}, {van Elteren},
  {Van Hemelryck}, {van Leeuwen}, {Vaschetto}, {Vecchiato}, {Veljanoski},
  {Viala}, {Vicente}, {Vogt}, {von Essen}, {Voss}, {Votruba}, {Voutsinas},
  {Walmsley}, {Weiler}, {Wertz}, {Wevers}, {Wyrzykowski}, {Yoldas},
  {{\v{Z}}erjal}, {Ziaeepour}, {Zorec}, {Zschocke}, {Zucker}, {Zurbach}, \&
  {Zwitter}}]{Mignard2018}
{Gaia Collaboration}, {Mignard}, F., {Klioner}, S.~A., {et~al.}
  2018{\natexlab{b}}, \aap, 616, A14, \dodoi{10.1051/0004-6361/201832916}

\bibitem[{{Graziani} {et~al.}(2019){Graziani}, {Courtois}, {Lavaux}, {Hoffman},
  {Tully}, {Copin}, \& {Pomar{\`e}de}}]{Graziani2019}
{Graziani}, R., {Courtois}, H.~M., {Lavaux}, G., {et~al.} 2019, \mnras, 488,
  5438, \dodoi{10.1093/mnras/stz078}

\bibitem[{{Gwinn} {et~al.}(1997){Gwinn}, {Eubanks}, {Pyne}, {Birkinshaw}, \&
  {Matsakis}}]{Gwinn1997}
{Gwinn}, C.~R., {Eubanks}, T.~M., {Pyne}, T., {Birkinshaw}, M., \& {Matsakis},
  D.~N. 1997, \apj, 485, 87, \dodoi{10.1086/304424}

\bibitem[{{Hall}(2019)}]{Hall2018}
{Hall}, A. 2019, \mnras, 486, 145, \dodoi{10.1093/mnras/stz648}

\bibitem[{{Hinshaw} {et~al.}(2009){Hinshaw}, {Weiland}, {Hill}, {Odegard},
  {Larson}, {Bennett}, {Dunkley}, {Gold}, {Greason}, {Jarosik}, {Komatsu},
  {Nolta}, {Page}, {Spergel}, {Wollack}, {Halpern}, {Kogut}, {Limon}, {Meyer},
  {Tucker}, \& {Wright}}]{Hinshaw2009}
{Hinshaw}, G., {Weiland}, J.~L., {Hill}, R.~S., {et~al.} 2009, \apjs, 180, 225,
  \dodoi{10.1088/0067-0049/180/2/225}

\bibitem[{{Hogg}(1999)}]{Hogg1999}
{Hogg}, D.~W. 1999, arXiv e-prints, astro.
\newblock \doarXiv{astro-ph/9905116}

\bibitem[{{Kardashev}(1986)}]{Kardashev1986}
{Kardashev}, N.~S. 1986, \azh, 63, 845

\bibitem[{{Lindegren} {et~al.}(2018){Lindegren}, {Hern{\'a}ndez}, {Bombrun},
  {Klioner}, {Bastian}, {Ramos-Lerate}, {de Torres}, {Steidelm{\"u}ller},
  {Stephenson}, {Hobbs}, {Lammers}, {Biermann}, {Geyer}, {Hilger}, {Michalik},
  {Stampa}, {McMillan}, {Casta{\~n}eda}, {Clotet}, {Comoretto}, {Davidson},
  {Fabricius}, {Gracia}, {Hambly}, {Hutton}, {Mora}, {Portell}, {van Leeuwen},
  {Abbas}, {Abreu}, {Altmann}, {Andrei}, {Anglada}, {Balaguer-N{\'u}{\~n}ez},
  {Barache}, {Becciani}, {Bertone}, {Bianchi}, {Bouquillon}, {Bourda},
  {Br{\"u}semeister}, {Bucciarelli}, {Busonero}, {Buzzi}, {Cancelliere},
  {Carlucci}, {Charlot}, {Cheek}, {Crosta}, {Crowley}, {de Bruijne}, {de
  Felice}, {Drimmel}, {Esquej}, {Fienga}, {Fraile}, {Gai}, {Garralda},
  {Gonz{\'a}lez-Vidal}, {Guerra}, {Hauser}, {Hofmann}, {Holl}, {Jordan},
  {Lattanzi}, {Lenhardt}, {Liao}, {Licata}, {Lister}, {L{\"o}ffler},
  {Marchant}, {Martin-Fleitas}, {Messineo}, {Mignard}, {Morbidelli}, {Poggio},
  {Riva}, {Rowell}, {Salguero}, {Sarasso}, {Sciacca}, {Siddiqui}, {Smart},
  {Spagna}, {Steele}, {Taris}, {Torra}, {van Elteren}, {van Reeven}, \&
  {Vecchiato}}]{Lindegren2018}
{Lindegren}, L., {Hern{\'a}ndez}, J., {Bombrun}, A., {et~al.} 2018, \aap, 616,
  A2, \dodoi{10.1051/0004-6361/201832727}

\bibitem[{{Mignard} \& {Klioner}(2012)}]{Mignard2012}
{Mignard}, F., \& {Klioner}, S. 2012, \aap, 547, A59,
  \dodoi{10.1051/0004-6361/201219927}

\bibitem[{{Paine} {et~al.}(2018){Paine}, {Darling}, \&
  {Truebenbach}}]{Paine2018}
{Paine}, J., {Darling}, J., \& {Truebenbach}, A. 2018, \apjs, 236, 37,
  \dodoi{10.3847/1538-4365/aabe2d}

\bibitem[{{Planck Collaboration} {et~al.}(2018){Planck Collaboration},
  {Aghanim}, {Akrami}, {Ashdown}, {Aumont}, {Baccigalupi}, {Ballardini},
  {Banday}, {Barreiro}, {Bartolo}, {Basak}, {Battye}, {Benabed}, {Bernard},
  {Bersanelli}, {Bielewicz}, {Bond}, {Borrill}, {Bouchet}, {Burigana},
  {Calabrese}, {Carron}, {Chiang}, {Comis}, {Contreras}, {Crill}, {Curto},
  {Cuttaia}, {de Bernardis}, {de Rosa}, {de Zotti}, {Delabrouille}, {Di
  Valentino}, {Dickinson}, {Diego}, {Dor{\'e}}, {Ducout}, {Dupac}, {Elsner},
  {En{\ss}lin}, {Eriksen}, {Falgarone}, {Fantaye}, {Finelli}, {Forastieri},
  {Frailis}, {Fraisse}, {Franceschi}, {Frolov}, {Galeotta}, {Galli}, {Ganga},
  {Gerbino}, {G{\'o}rski}, {Gruppuso}, {Gudmundsson}, {Hand ley}, {Hansen},
  {Herranz}, {Hivon}, {Huang}, {Jaffe}, {Keih{\"a}nen}, {Keskitalo}, {Kiiveri},
  {Kim}, {Kisner}, {Krachmalnicoff}, {Kunz}, {Kurki-Suonio}, {Lamarre},
  {Lasenby}, {Lattanzi}, {Lawrence}, {Le Jeune}, {Levrier}, {Liguori}, {Lilje},
  {Lindholm}, {L{\'o}pez-Caniego}, {Lubin}, {Ma}, {Mac{\'\i}as-P{\'e}rez},
  {Maggio}, {Maino}, {Mand olesi}, {Mangilli}, {Martin},
  {Mart{\'\i}nez-Gonz{\'a}lez}, {Matarrese}, {Mauri}, {McEwen}, {Melchiorri},
  {Mennella}, {Migliaccio}, {Miville-Desch{\^e}nes}, {Molinari}, {Moneti},
  {Montier}, {Morgante}, {Natoli}, {Oxborrow}, {Pagano}, {Paoletti},
  {Partridge}, {Perdereau}, {Perotto}, {Pettorino}, {Piacentini},
  {Plaszczynski}, {Polastri}, {Polenta}, {Rachen}, {Racine}, {Reinecke},
  {Remazeilles}, {Renzi}, {Rocha}, {Roudier}, {Ruiz-Granados}, {Sandri},
  {Savelainen}, {Scott}, {Sirignano}, {Sirri}, {Spencer}, {Stanco}, {Sunyaev},
  {Tauber}, {Tavagnacco}, {Tenti}, {Toffolatti}, {Tomasi}, {Tristram},
  {Trombetti}, {Valiviita}, {Van Tent}, {Vielva}, {Villa}, {Vittorio},
  {Wandelt}, {Wehus}, {Zacchei}, \& {Zonca}}]{Aghanim2018}
{Planck Collaboration}, {Aghanim}, N., {Akrami}, Y., {et~al.} 2018, \aap, 617,
  A48, \dodoi{10.1051/0004-6361/201731489}

\bibitem[{{Taylor}(2005)}]{topcat}
{Taylor}, M.~B. 2005, Astronomical Society of the Pacific Conference Series,
  Vol. 347, {TOPCAT \&amp; STIL: Starlink Table/VOTable Processing Software},
  ed. P.~{Shopbell}, M.~{Britton}, \& R.~{Ebert}, 29

\bibitem[{{Titov} \& {Lambert}(2013)}]{Titov2013}
{Titov}, O., \& {Lambert}, S. 2013, \aap, 559, A95,
  \dodoi{10.1051/0004-6361/201321806}

\bibitem[{{Truebenbach} \& {Darling}(2017)}]{Truebenbach2017}
{Truebenbach}, A.~E., \& {Darling}, J. 2017, \apjs, 233, 3,
  \dodoi{10.3847/1538-4365/aa9026}

\bibitem[{{Tully} {et~al.}(2016){Tully}, {Courtois}, \& {Sorce}}]{Tully2016}
{Tully}, R.~B., {Courtois}, H.~M., \& {Sorce}, J.~G. 2016, \aj, 152, 50,
  \dodoi{10.3847/0004-6256/152/2/50}

\bibitem[{{Tully} \& {Fisher}(1977)}]{Tully1977}
{Tully}, R.~B., \& {Fisher}, J.~R. 1977, \aap, 500, 105

\end{thebibliography}

\texttt{\appendix}
\section{Galaxies}\label{app:images}

Galaxies in the {\it Gaia}-Cosmicflows crossmatch were each
visually inspected for features that would indicate a poor {\it Gaia} fit, as discussed 
in Section~\ref{sec:visual inspection}. 
Figures \ref{fig:images1}-\ref{fig:images8} show images of the 
232 galaxies used for the parallax limit in Section~\ref{sec:limit}. SDSS $g$-band images
are shown where available, otherwise the images are DSS2 red-band. 
Several galaxies are located near the edge of the SDSS plate, in which case we 
reviewed additional images in the literature, but we display the truncated SDSS images in 
Figures \ref{fig:images1}-\ref{fig:images8}.

\begin{figure*}[hbt!]
    \centering
    \includegraphics[width=\textwidth]{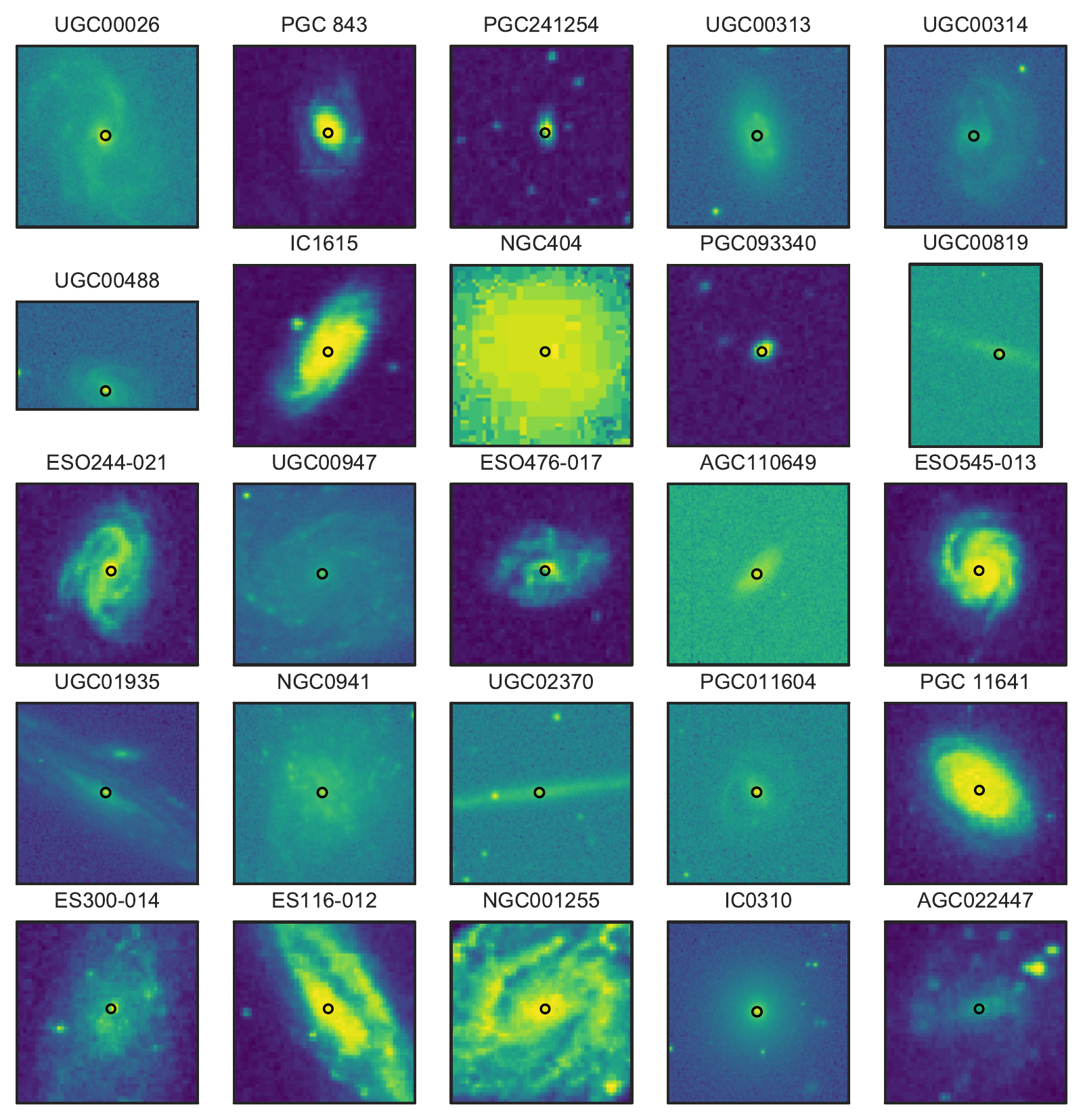}
    \caption{SDSS g-band or DSS2 red-band imaging for galaxies selected to attempt detection of the secular parallax dipole.}
    \label{fig:images1}
\end{figure*}

\begin{figure*}[hbt!]
    \centering
    \includegraphics[width=\textwidth]{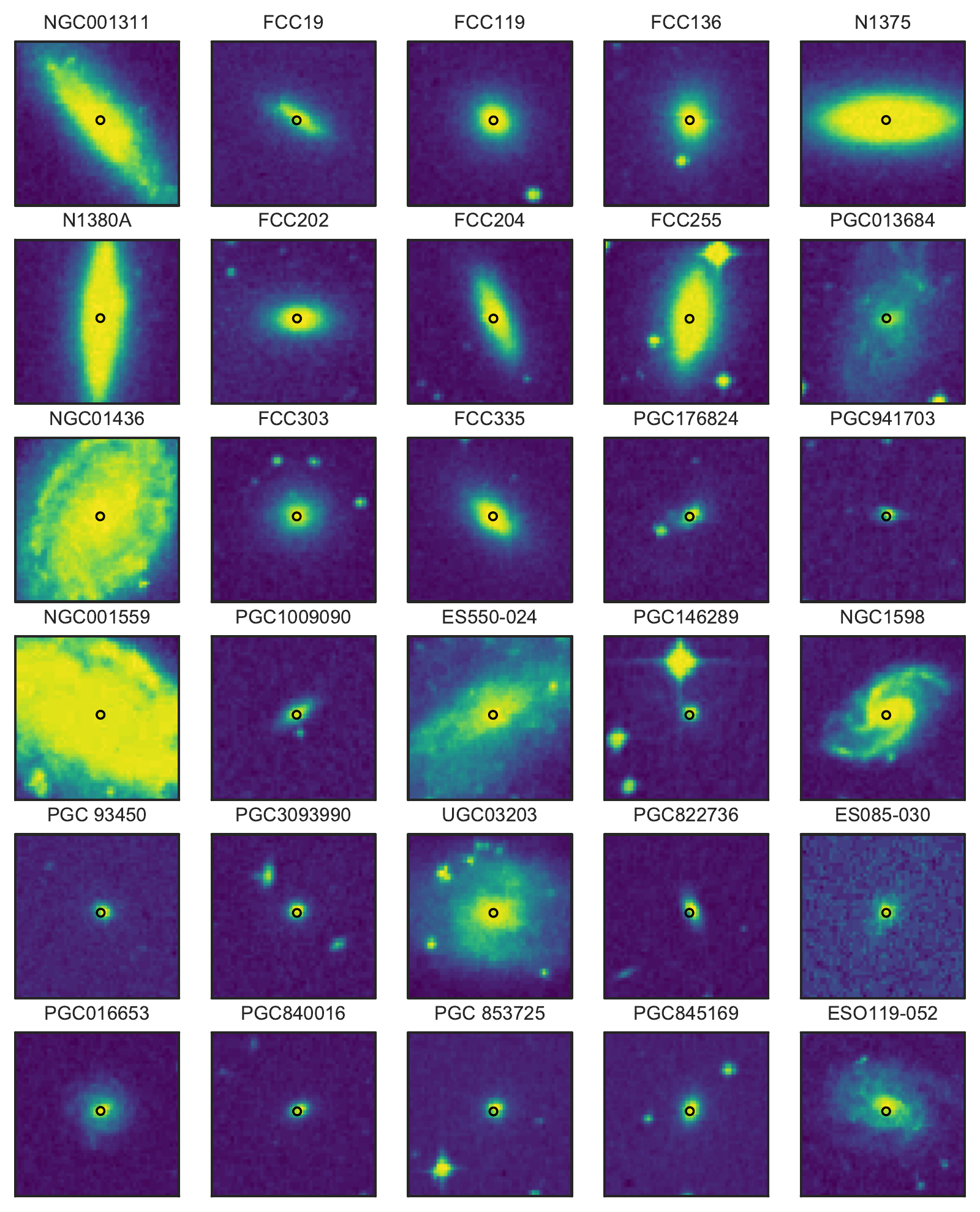}
    \caption{Continued from Figure~\ref{fig:images1}.}
    \label{fig:images2}
\end{figure*}

\begin{figure*}[hbt!]
    \centering
    \includegraphics[width=\textwidth]{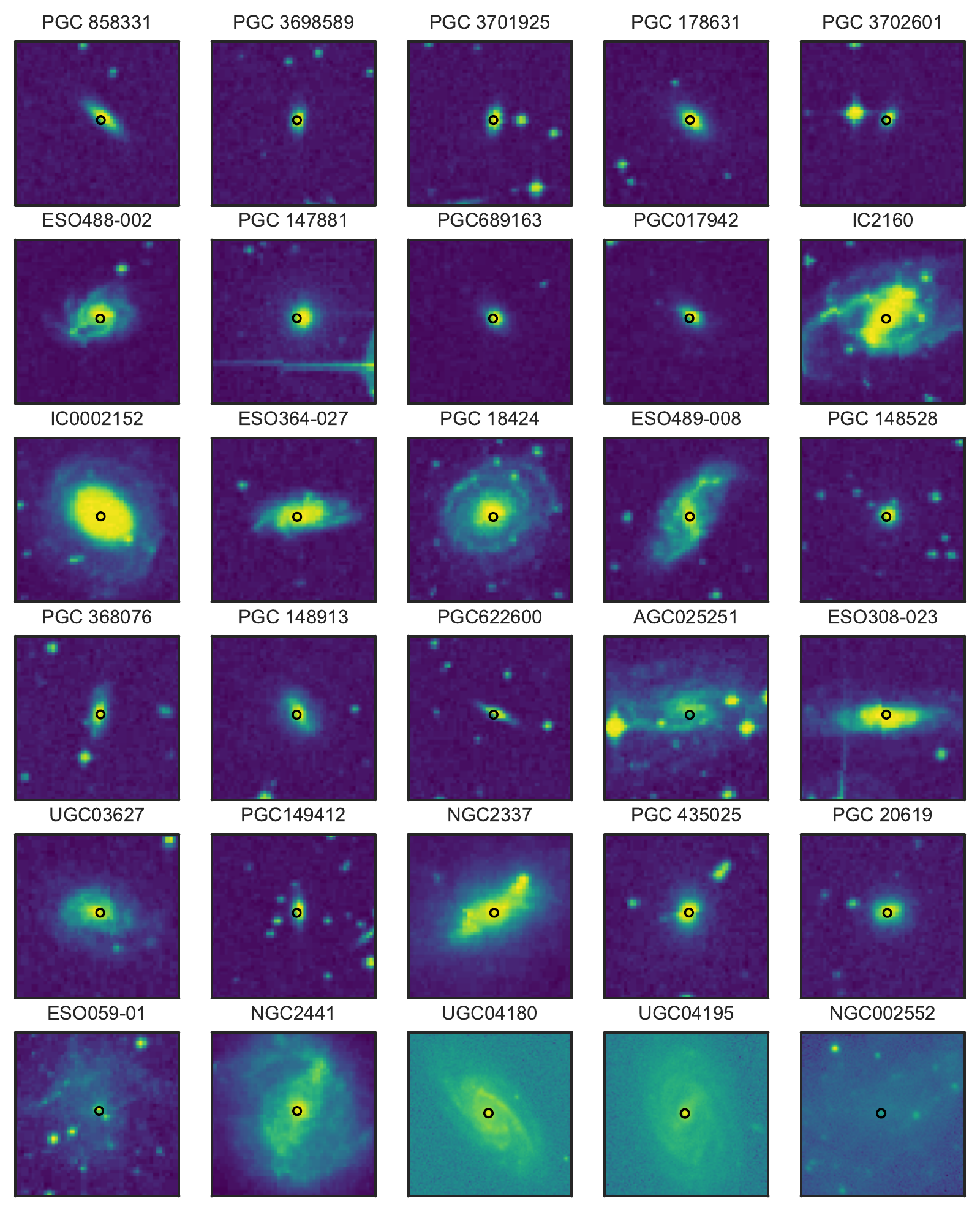}
    \caption{Continued from Figure~\ref{fig:images2}.}
    \label{fig:images3}
\end{figure*}

\begin{figure*}[hbt!]
    \centering
    \includegraphics[width=\textwidth]{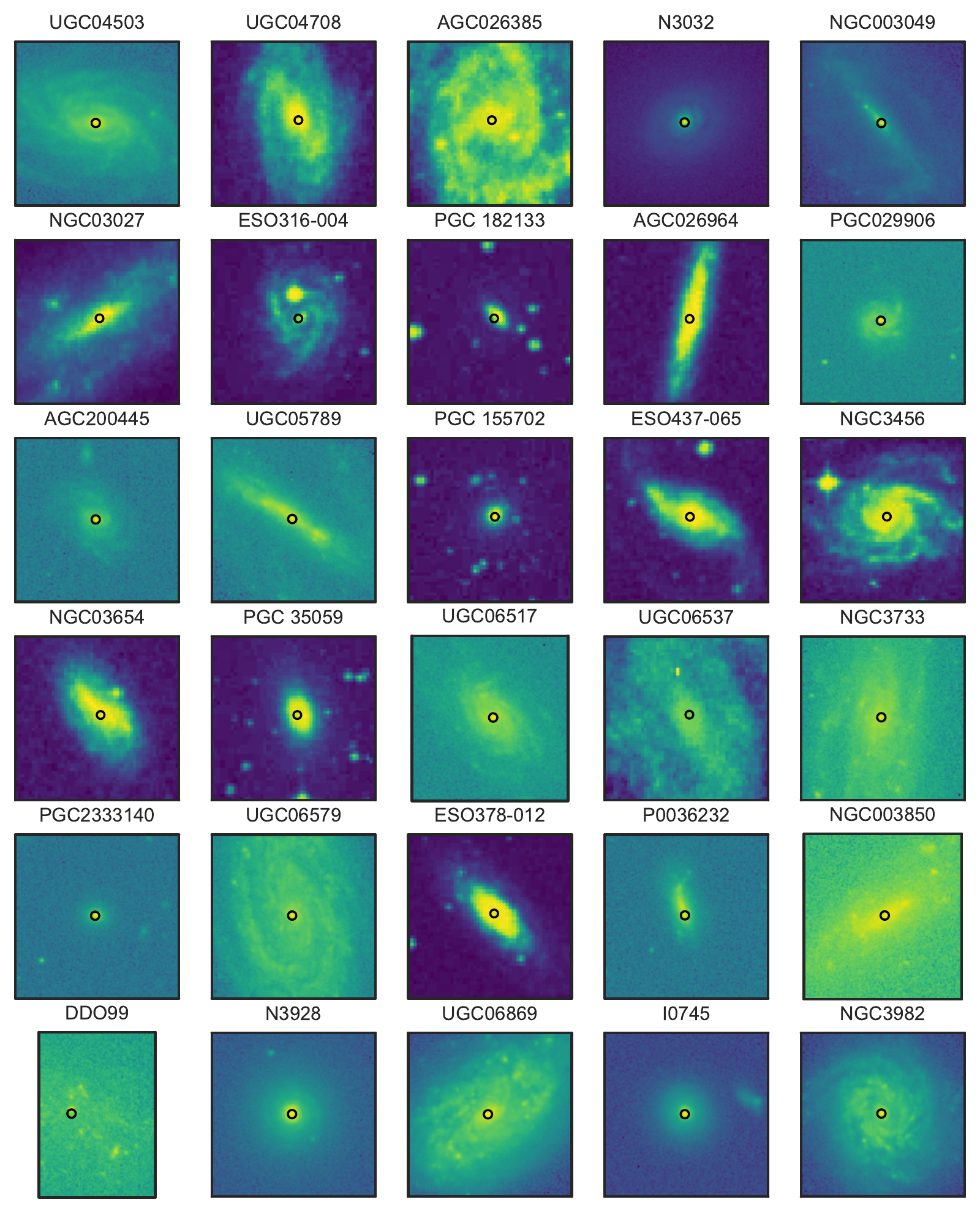}
    \caption{Continued from Figure~\ref{fig:images3}.}
    \label{fig:images4}
\end{figure*}

\begin{figure*}[hbt!]
    \centering
    \includegraphics[width=\textwidth]{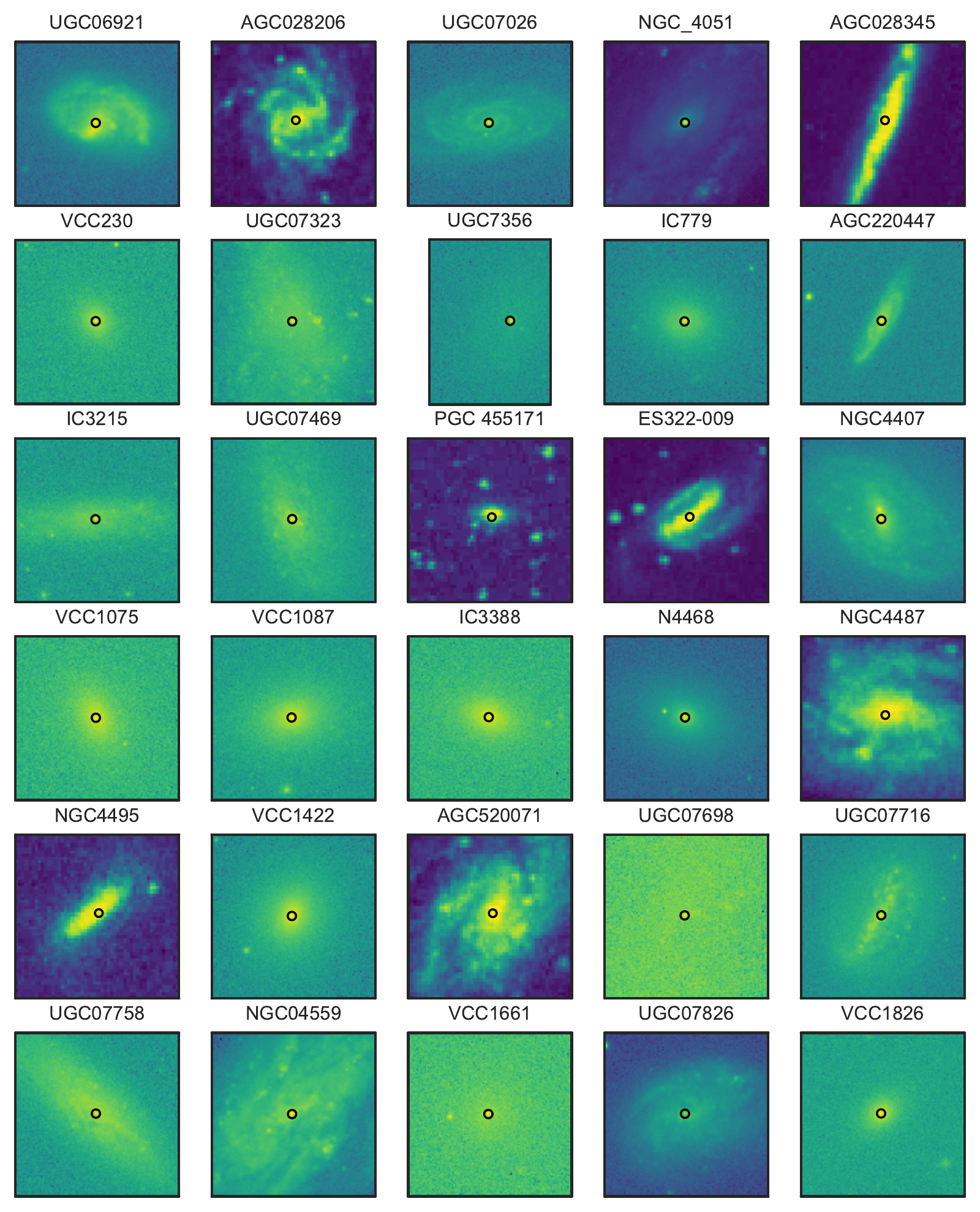}
    \caption{Continued from Figure~\ref{fig:images4}.}
    \label{fig:images5}
\end{figure*}

\begin{figure*}[hbt!]
    \centering
    \includegraphics[width=\textwidth]{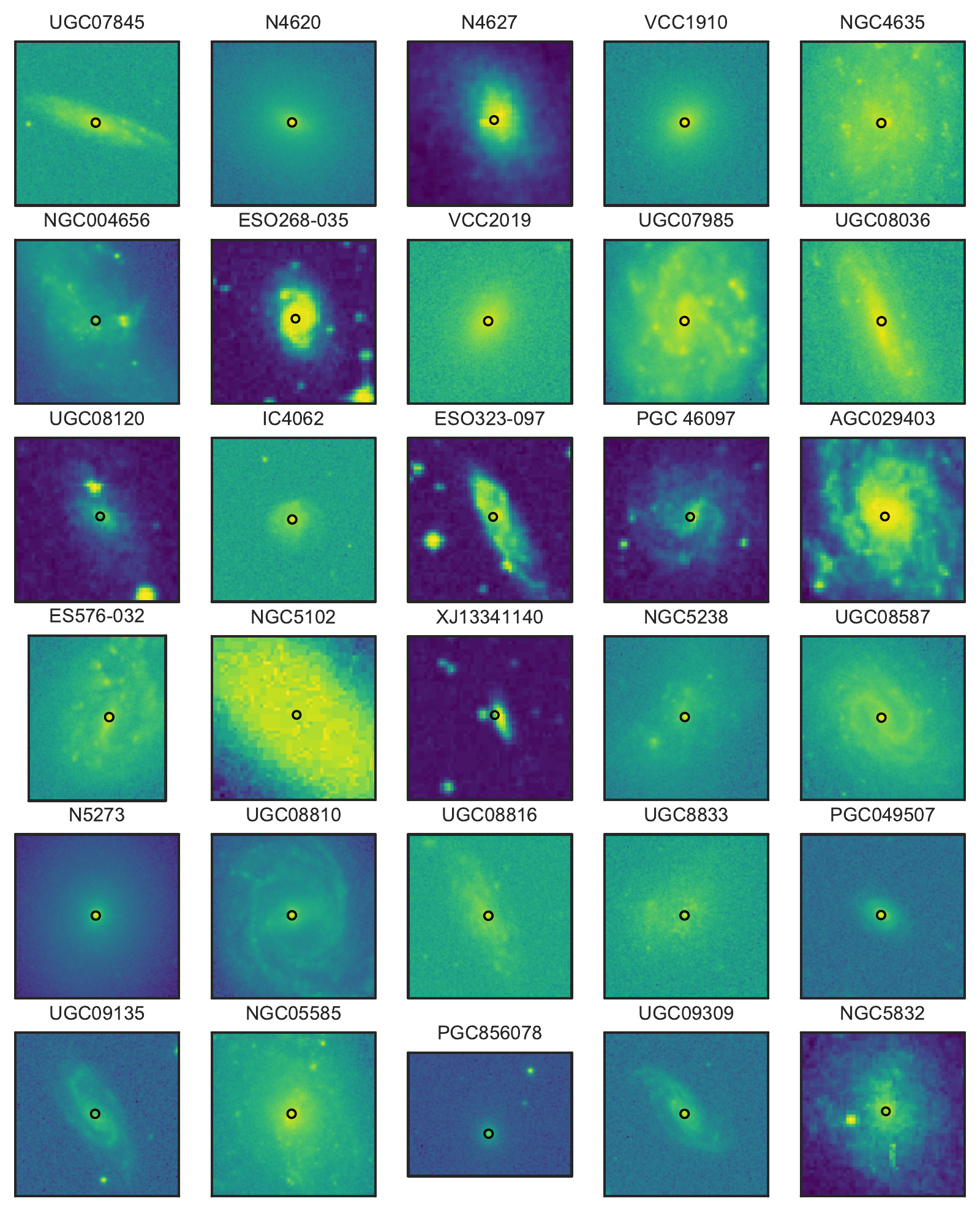}
    \caption{Continued from Figure~\ref{fig:images5}.}
    \label{fig:images6}
\end{figure*}

\begin{figure*}[hbt!]
    \centering
    \includegraphics[width=\textwidth]{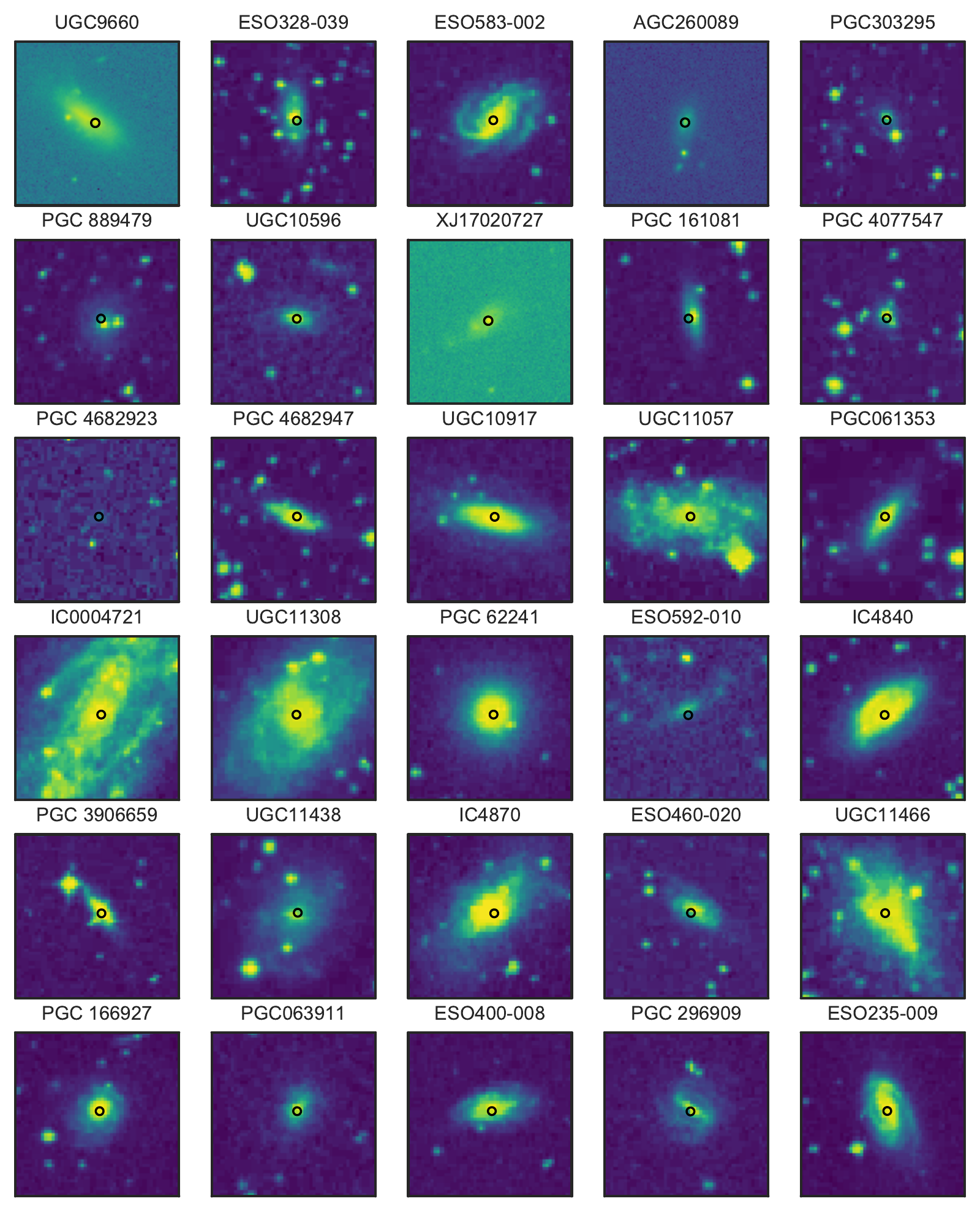}
    \caption{Continued from Figure~\ref{fig:images6}.}
    \label{fig:images7}
\end{figure*}

\begin{figure*}[hbt!]
    \centering
    \includegraphics[width=\textwidth]{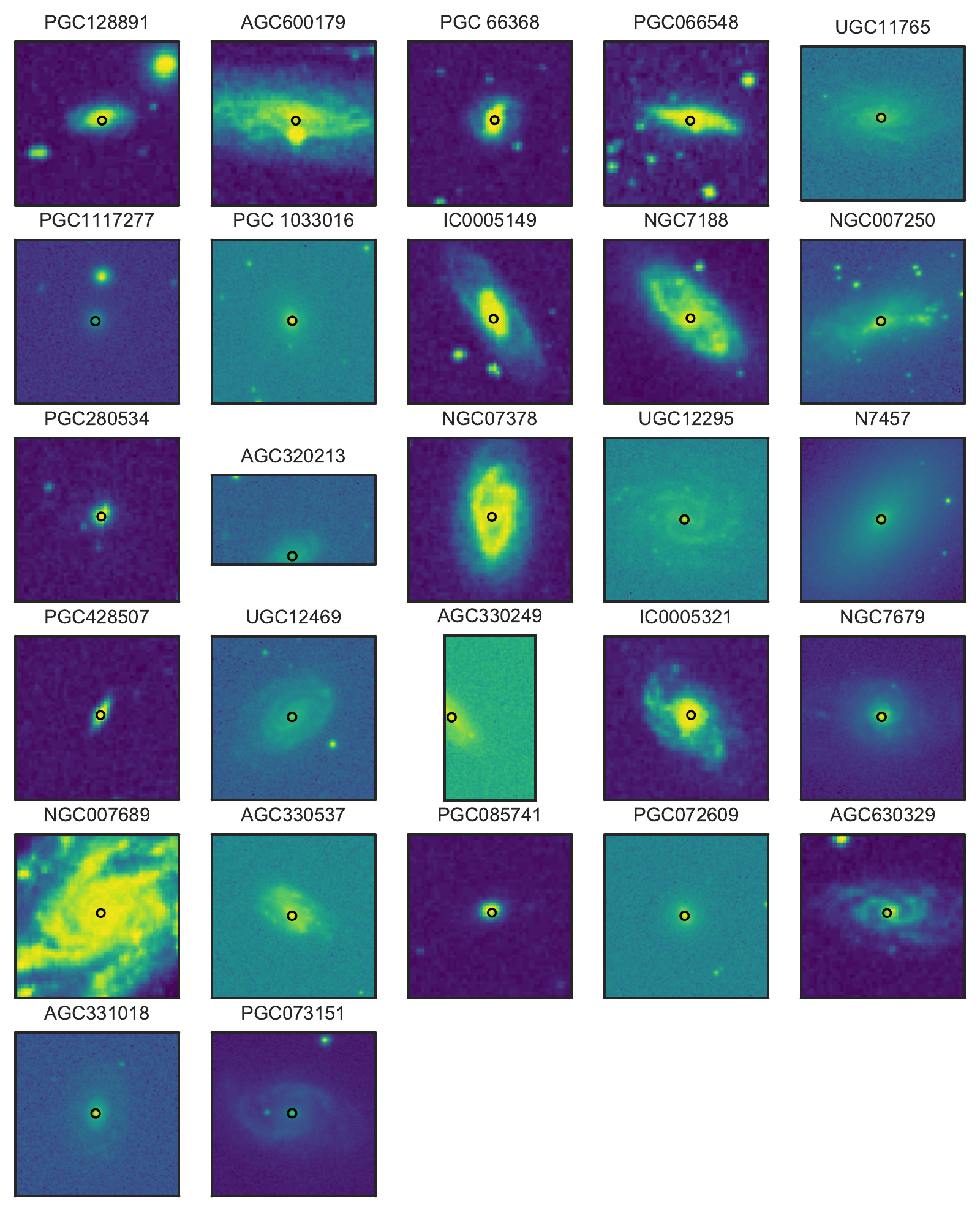}
    \caption{Continued from Figure~\ref{fig:images7}.}
    \label{fig:images8}
\end{figure*}

\end{document}